\documentclass[twocolumn,reprint,amsmath,amssymb,footinbib,superscriptaddress,floatfix,aps,longbibliography,prm]{revtex4-2}
\usepackage{graphicx}
\usepackage{amsmath}
\usepackage{braket}
\usepackage{amssymb}
\usepackage{siunitx}
\usepackage{lineno,hyperref}
\usepackage{tabularx}
\usepackage{float}

\newcommand{\Ns}[1]{$^{#1}{\mathrm{N}_{\mathrm{s}}}^0$}
\newcommand{\NVH}[1]{$^{#1}\mathrm{NVH}^{-}$}

\begin{document}

\title{Detecting nitrogen-vacancy-hydrogen centers on the nanoscale using nitrogen-vacancy centers in diamond} 

\author{Christoph Findler}
\email[]{christoph.findler@uni-ulm.de}
\affiliation{Institute for Quantum Optics, Ulm University, Albert-Einstein-Allee 11, D-89081 Ulm, Germany}
\affiliation{Diatope GmbH, Buchenweg 23, D-88444 Ummendorf, Germany}
\author{Rémi Blinder}

\author{Karolina Schüle}

\affiliation{Institute for Quantum Optics, Ulm University, Albert-Einstein-Allee 11, D-89081 Ulm, Germany}

\author{Priyadharshini Balasubramanian}
\affiliation{Institute for Quantum Optics, Ulm University, Albert-Einstein-Allee 11, D-89081 Ulm, Germany}

\author{Christian Osterkamp}
\affiliation{Diatope GmbH, Buchenweg 23, D-88444 Ummendorf, Germany}
\affiliation{Institute for Quantum Optics, Ulm University, Albert-Einstein-Allee 11, D-89081 Ulm, Germany}

\author{Fedor Jelezko}
\affiliation{Institute for Quantum Optics, Ulm University, Albert-Einstein-Allee 11, D-89081 Ulm, Germany}

\date{\today}

\begin{abstract}
In diamond, nitrogen defects like the substitutional nitrogen defect (N$_\mathrm{s}$) or the nitrogen-vacancy-hydrogen complex (NVH) outnumber the nitrogen vacancy (NV) defect by at least one order of magnitude creating a dense spin bath. While neutral N$_\mathrm{s}$ has an impact on the coherence of the NV spin state, the atomic structure of NVH reminds of a NV center decorated with a hydrogen atom. As a consequence, the formation of NVH centers could compete with that of NV centers possibly lowering the N-to-NV conversion efficiency in diamond grown with hydrogen-plasma-assisted chemical vapor deposition (CVD). Therefore, monitoring and controlling the spin bath is essential to produce and understand engineered diamond material with high NV concentrations for quantum applications. While the incorporation of N$_\mathrm{s}$ in diamond has been investigated on the nano- and mesoscale for years, studies concerning the influence of CVD parameters and the crystal orientation on the NVH formation have been restricted to bulk N-doped diamond providing high-enough spin numbers for electron paramagnetic resonance and optical absorption spectroscopy techniques. 
Here, we investigate sub-micron-thick (100)-diamond layers with nitrogen contents of (13.8$\,\pm\,$1.6)$\,$ppm and (16.7$\,\pm\,$3.6)$\,$ppm, and exploiting the NV centers in the layers as local nano-sensors, we demonstrate the detection of \NVH{} centers using double-electron-electron-resonance (DEER). To determine the \NVH{} densities, we quantitatively fit the hyperfine structure of \NVH{} and confirm the results with the DEER method usually used for determining \Ns{} densities. With our experiments, we access the spin bath composition on the nanoscale and enable a fast feedback-loop in CVD recipe optimization with thin diamond layers instead of resource- and time-intensive bulk crystals. Furthermore, the quantification of \NVH{} plays a very important role for understanding the dynamics of vacancies and the incorporation of hydrogen into CVD diamond optimized for quantum technologies.
\end{abstract}


\maketitle 

\section{Introduction}

Solid-state quantum systems based on point defects in crystals like silicon carbide and diamond are promising platforms for quantum technologies since they can host fluorescent electron spin defects with long coherence times\cite{Christle.2015,Balasubramanian.2009,Awschalom.2018}. Typical defect concentrations range from parts-per-trillion (ppt) for single and isolated centers to dense ensembles with several parts-per-million (ppm) defects in the host material. As a consequence, analysing these defects and their local environment is challenging, in particular for ultra-pure crystals. In case of fluorescent defects, confocal microscopy is capable of resolving single atomic defects, like, for instance, the negatively charged nitrogen vacancy center (NV$^-$) in diamond\cite{Jelezko.2004}. The defect consists of a substitutional nitrogen, an adjacent carbon vacancy and an electron from a donor (Fig.~\ref{Fig1}a). Its electron spin ($S = 1$) exhibits coherence times up to several milliseconds\cite{Balasubramanian.2009,Herbschleb.2019} and it can be controlled by laser and microwave pulses\cite{Jelezko.2004}. Its atomic size and the high sensitivity to magnetic fields\cite{Schmitt.2017,Vetter.2022} makes the NV$^-$ an ideal sensor for nuclear magnetic resonance spectroscopy in nano-\cite{Staudacher.2013,Staudenmaier.2022,Liu.2022} and micron-sized volumes\cite{Glenn.2018,Bucher.2020}. Likewise, the sensing characteristics of the NV$^-$ center can also be exploited to analyze the density and nature of electron spins\cite{Stepanov.2016} that usually form when doping diamond with nitrogen\cite{Edmonds.2012}.

Hydrogen-plasma-assisted chemical-vapor-deposition (CVD) of diamond represents currently the state-of-the-art method to synthesize N-doped diamond for NV$^-$ applications with adjustable thickness\cite{Ohno.2012b, Eichhorn.2019,ChristianOsterkamp.2015} and concentration\cite{Lobaev.2017,Osterkamp.2020,Edmonds.2021}. Despite of the progress in engineering CVD-grown diamond, low N-to-NV conversion efficiencies of a few percent or less\cite{Edmonds.2012,Osterkamp.2020,Tallaire.2020,Edmonds.2021,Balasubramanian.2022} remain still a problem. In other words, the CVD recipes of today produce mostly other nitrogen-related defects than NV$^-$.

In (100)-oriented diamond, for example, substitutional nitrogen (N$_\mathrm{s}$) can be up to 300-times more abundant than NV$^-$\cite{Edmonds.2012}. Substitutional nitrogen in its neutral charge state (\Ns{} or P1) is a paramagnetic but non-fluorescent (dark) defect carrying a $S=1/2$ spin\cite{Cox.1994} and, hence, constitutes a spin bath around the NV$^-$ centers (Fig.~\ref{Fig1}b) often limiting their coherence time $T_2$\cite{Bauch.2020}. As a consequence, a high NV$^-$:\Ns{} ratio is favorable for sensing applications as it results in reduced magnetic noise, i.e. enhanced coherence time, and an increased number of interrogated NV$^-$ centers, potentially leading to higher signal-to-noise ratio\cite{Rondin.2014}.

One way for increasing the NV$^-$ concentration in CVD diamond is generating additional vacancies by electron irradiation and converting them together with \Ns{} by annealing into NV$^-$  \cite{Edmonds.2021,Luo.2022}. Other studies optimize the CVD process itself by varying the substrate orientation\cite{Osterkamp.2020,Balasubramanian.2022}, the nitrogen source\cite{Tallaire.2020}, the N:C ratio in the gas phase\cite{Osterkamp.2020,Luo.2022}, or the growth temperature\cite{Chouaieb.2019,Balasubramanian.2022}. In particular, changing from the growth on (100)- to (113)-oriented substrates and optimizing the growth temperature has been reported to yield NV$^-$:\Ns{} ratios of up to 25$\%$ without any post-treatment\cite{Balasubramanian.2022}. The origin of the improvement remains unclear and a better understanding of the generation of defects in CVD-grown diamond is thus crucial. 

The second most abundant spin species in CVD-diamond\cite{Edmonds.2012}, the negatively charged nitrogen vacancy hydrogen defect (NVH$^-$), structurally reminds of a NV center decorated with an additional hydrogen atom\cite{Glover.2003}(Fig.~1b). The latter undergoes fast tunneling between the carbons adjacent to the defect's vacancy causing an effective [111]-oriented $C_{3v}$ symmetry\cite{Glover.2003,Goss.2003}, similar to that of the NV center (Fig.~1b). As a consequence, being usually one order of magnitude more abundant than NV$^-$ in (100)-oriented CVD-diamond\cite{Edmonds.2012,Edmonds.2021}, the NVH defect comes under suspicion to reduce the N-to-NV conversion degree. The formation mechanism of NVH remains still unknown, though. One possibility is the incorporation as whole units\cite{Edmonds.2012} and/or NV centers get passivated after their formation by H-atoms\cite{Edmonds.2012,Stacey.2012,Findler.2020}. Either way, NVH centers have to be considered as a possible sink for NV centers which is why their density should be studied carefully and their generation understood to develop superior diamond growth.

The main technique to investigate paramagnetic spins is electron paramagnetic resonance (EPR) spectroscopy\cite{Eaton.2010}. For bulk single-crystalline diamond with a sample volume of a few mm$^{3}$, conventional EPR spectrometers are sensitive to defect concentrations down to parts-per-billion (ppb)\cite{Edmonds.2012}. This capability has proven useful in several cases of in-depth studies on bulk diamond material\cite{Edmonds.2012,Edmonds.2021,Luo.2022} or defect dynamics induced by thermal annealing\cite{Lomer.1973,D.J.Twitchen.2001,Yamamoto.2013}. However, for quantum applications on the nano-/ mesoscale like nuclear magnetic resonance spectroscopy\cite{Staudacher.2013,Glenn.2018,Bucher.2020} or wide-field magnetometry of substances outside of the diamond\cite{Pham.2011,Glenn.2015,Simpson.2016,Tetienne.2017,Healey.2020}, single centers or thin NV$^-$-layers with thicknesses of a few microns close to the surface are required. As a result, the number of spins involved lies several orders of magnitude below the sensitivity of conventional EPR. Likewise, techniques based on absorption spectroscopy (IR or UV-visible)\cite{Tallaire.2020, Luo_uv_ns0.2022} show equivalent sensitivity limits for thin layers.

Optically detected double-electron-electron-resonance (DEER) using NV$^-$ centers as local magnetic field sensors, on the other hand, works even with single atomic centers\cite{Degen.2021} and the technique has already been applied to investigate \Ns{} in nm-thin layers\cite{Eichhorn.2019,Osterkamp.2020,Li.2021,Balasubramanian.2022}. In this work, we demonstrate that DEER is also capable of detecting NVH$^-$, even for densities of only a few hundreds of ppb. The defect nature is confirmed by resolving the hyperfine structure expected for NVH$^-$ and we determine the bath density by quantitatively fitting DEER spectra of sub-micron-thick layers considering multiple spin species and forbidden transitions. Finally, the obtained NVH$^-$ concentrations are verified by following a well-established DEER approach for estimating \Ns{} bath densities\cite{Stepanov.2016,Eichhorn.2019,Osterkamp.2020,Balasubramanian.2022}.

\section{Results and Discussion}
For detecting NVH$^{-}$ in thin diamond layers, we prepare two $^{15}$N-doped samples with slightly different nitrogen concentrations as determined by secondary-ion-mass-spectrometry (SIMS), i.e. (13.8$\,\pm\,$1.6)$\,$ppm for sample A and (16.7$\,\pm\,$3.6)$\,$ppm for sample B. According to SIMS, the thicknesses of the N-doped layers are respectively (570$\,\pm\,$60)$\,$nm and (700$\,\pm\,$50)$\,$nm for sample A and B. Both layers are homoepitaxially grown on (100)-oriented single-crystalline substrates using microwave-assisted plasma chemical vapor deposition (CVD) with $^{12}$C-enriched (99.99$\,\%$) methane and $^{15}$N$_2$ (98$\,\%$) gas. The details of the CVD process are provided in the Supplemental Material\cite{Supinfo1.1}.

\begin{figure*} [ht]
	\begin{center}
	\includegraphics[scale=1.0]{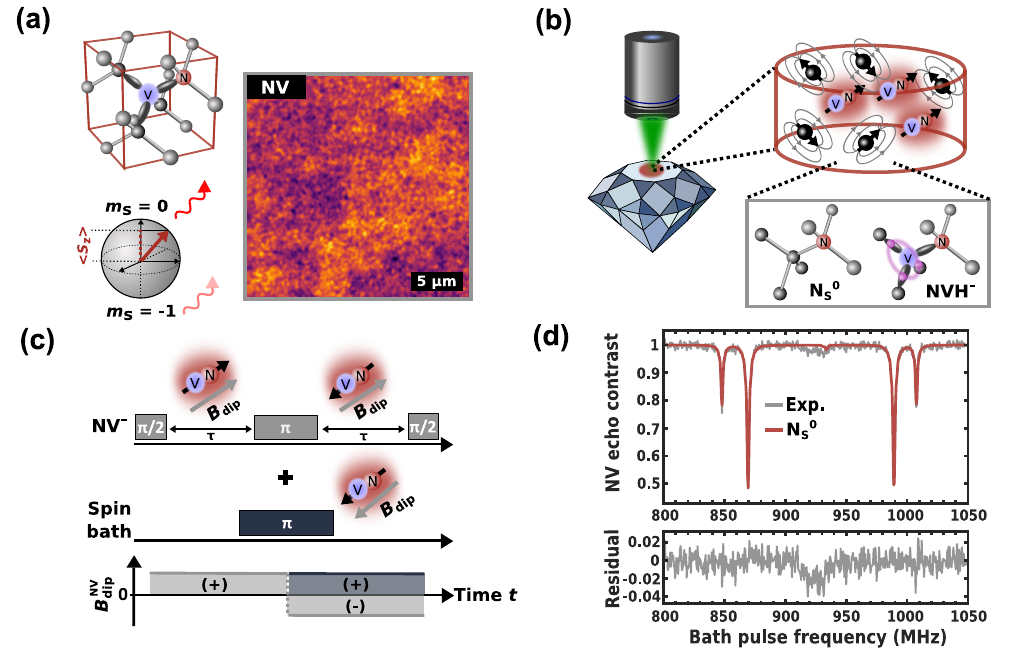}
	\end{center}
	\caption{a) The structure of the NV-center in diamond and a confocal microscopy image of sample A. The Bloch-sphere illustrates the spin-dependent fluorescence that is exploited for spin-state read-out. b) Confocal microscopy analysis of a NV$^-$ ensemble and the local paramagnetic spin bath consisting of substitutional nitrogen (\Ns{}) and the negatively-charged nitrogen-vacancy-hydrogen complex (\NVH{}). The quasi-static magnetic field is measured through the NV$^{-}$ ensemble  c) Double-electron-electron-resonance (DEER) pulse sequence for probing the quasi-static, dipolar magnetic field $B_{dip}$ of paramagnetic spins using NV$^-$ centers as local sensors. d) Experimental DEER spectrum for sample A at 330$\,$G aligned along $\left< 111 \right>$ including a fit with a \Ns{15}-spin model. The bath $\pi$-pulse is 525$\,$ns long and $\tau$=1.5$\,\mathrm{\mu}$s.}
	\label{Fig1}
	\end{figure*}

Utilizing a confocal microscope, we probe the fluorescence of NV$^{-}$ centers by optical excitation with a green laser. By scanning the laser over the diamond we obtain a confocal microscopy image as shown in Fig.~\ref{Fig1}a (sample A). In total we investigate three NV$^{-}$ ensembles per diamond sample labelled A1,2,3 and B1,2,3, respectively. The analysis of the dark spin environment is first demonstrated using ensemble A1 and B1 as an example and later extended to the remaining ensembles.

Considering the thickness of the doped layers and the dimensions of our confocal volume (300$\,$x$\,$300$\,$x$\,$500$\,$nm$^{3}$ (FWHM)), the layers studied in this work are completely illuminated by the laser and lie entirely in the confocal volume with respect to depth, i.e. can be considered as optically two-dimensional. With microwave pulses we control the spin state of the NV$^{-}$ centers in the interrogated volume by magnetic resonance. Dark, paramagnetic defects like \Ns{} and NVH$^{-}$, can also be addressed by microwaves but, due to the lack of spin-dependent fluorescence, no direct optical readout of their spin state has been reported at room temperature. The defect \Ns{}, however, causes a quasi-static magnetic field that can be sensed by NV$^{-}$ centers scattered around (Fig.~\ref{Fig1}b)\cite{Stepanov.2016,Eichhorn.2019}. Therefore, one can exploit the dipolar coupling between the fluorescent NV$^{-}$ centers and the dark spins to estimate the density of the latter ones\cite{Stepanov.2016,Eichhorn.2019,Osterkamp.2020}. For this we employ optically detected three-pulse DEER\cite{Stepanov.2016,Eichhorn.2019} where the first pulses on NV$^{-}$ ($\pi/2-\pi$) correspond to the usual Hahn-echo, and the last $\pi/2$ converts the magnetization from the equatorial plane to the $z$-direction for optical read-out (Fig.~\ref{Fig1}c). Without any $\pi$-pulse on the bath, the Hahn-echo decouples the NV$^{-}$ ensemble from the quasi-static, local magnetic field created by the bath spins $B_{dip}$ \cite{Stepanov.2016, Eichhorn.2019}. The corresponding phase accumulated for the NV$^{-}$-spin state on the Bloch sphere during the Hahn-echo sequence reads at the time $t=2\tau$ as \mbox{$\varphi_{2\tau} = \int_{t=0}^{2 \tau} B^{\mathrm{NV}}_{dip} \mathrm{d}t$}, with $B^{\mathrm{NV}}_{dip}=(B_{dip},-B_{dip})$ in the first ($t<\tau$) and second part of the sequence respectively. 
Consequently, the phases accumulated in the two  segments, [$0,\tau$]  and [$\tau, 2\tau$], 
of the Hahn-echo cancel each other (Fig.~\ref{Fig1}c). Adding the resonant bath $\pi$-pulse, however, leads to the simultaneous inversion of $B_{dip}$ with the NV$^{-}$ spin state resulting in a non-zero phase $\varphi_{2\tau} = \int_{t=0}^{2 \tau} B^{\mathrm{NV}}_{dip} \mathrm{d}t \neq 0$, which translates into decoherence (Fig.~\ref{Fig1}c).
The coherence left after re-coupling $B_{dip}$ of a spin bath with the density $n_{\mathrm{bath}}$ to the NV$^{-}$ ensemble is called DEER intensity and can be written as:\cite{Stepanov.2016,Eichhorn.2019}

\begin{equation}
C_{\mathrm{DEER}} =  C_{\tau} \exp\left(-A \gamma_{\mathrm{NV}} \gamma  \ n_{\mathrm{bath}} T \left< \sin^2\left(\frac{\theta}{2}\right) \right>_L \right),
\label{eq_contrast_1_compact}
\end{equation}
\noindent with $C_{\tau} = C_0\exp\left( -2\tau/T_2\right)$, $A=2\pi\mu_0 \hbar/9\sqrt{3}$, $T$ as the timing of the bath $\pi$-pulse relative to the NV$^{-}$ $\pi/2$-pulse (Fig.~\ref{Fig3}b), and  $\gamma_{\mathrm{NV}} = g_{\mathrm{NV}} \mu_B/\hbar$, $ \gamma = g \mu_B/\hbar$ as the gyromagnetic ratios for NV$^{-}$ and the bath spins, respectively. The constant $\hbar$ is the reduced Plank constant, $\mu_0$ the vacuum permeability, $\mu_B$ the Bohr magneton, $g_{\mathrm{NV}}$ and $ g$ are the $g$-factors of the NV$^{-}$ centers and the bath spins, respectively. The factor $C_0$ in $C_{\tau}$ represents the maximal contrast obtained from the NV$^{-}$ centers and the term $\exp\left( -2\tau/T_2\right)$ accounts for the loss of coherence during the DEER sequence ($t\,=\,2\tau$) that is not related to the partial inversion of the bath spins. The parameter $\theta$ is the flip angle induced by the pulse on the bath spins\cite{Stepanov.2016}. It is worth remarking that $\sin^2\left(\theta/2\right)$ in Eq.\ref{eq_contrast_1_compact} corresponds to the probability of spin inversion by the applied $\pi$-pulse. Thus, for a dark spin ensemble, the sine-term in Eq.\ref{eq_contrast_1_compact} represents the fraction of spins that are successfully inverted. In the ideal case ($\theta=\pi$) the inverted fraction is 1. Practically, however, proper inversion is attained only for spin species that are well on resonance with the pulse carrier frequency. To consider the actual distribution of flip angles across the dark spins, averaging of $\sin^2\left(\theta/2\right)$ needs to be performed over the full, inhomogeneously broadened spectrum denoted by the brackets $\left< \right>_L$ in Eq.\ref{eq_contrast_1_compact}.

For the DEER spectrum measured for NV$^{-}$ ensemble A1 in Fig.~\ref{Fig1}d, we fix $T\sim\tau$ in the Hahn-Echo and vary the frequency of the bath $\pi$-pulse around the transition frequency of a free $S=1/2$ electron spin in a static, external magnetic field $B_0$$\,$=$\,$330$\,$G, here $\nu_{\mathrm{free}}$$\,$=$\,$924$\,$MHz. Subtracting the DEER intensity obtained with a $3\pi/2$-pulse before the optical readout in Fig.~\ref{Fig1}c from the results with a $\pi/2$-pulse instead yields the NV echo contrast shown in Fig.~\ref{Fig1}d. We observe four major and one broad, small signals around $\nu_{\mathrm{free}}$ in the DEER spectrum (Fig.~\ref{Fig1}d). Since \Ns{15} is expected to be the most abundant paramagnetic defect in $^{15}$N-doped diamond\cite{Edmonds.2012,Stepanov.2016, Eichhorn.2019,Osterkamp.2020}, we fit Eq.~\ref{eq_contrast_1_compact} to the experimental data in (Fig.~\ref{Fig1}d) using the MATLAB toolbox Easyspin\cite{Stoll.2006}, the actual duration of the bath $\pi$-pulse, and the spin Hamiltonian for \Ns{15}, $\hat{H}=g\mu_B\vec{B}_{0}\hat{S}+\hat{S}\stackrel{\leftrightarrow}{A}\hat{I}$\cite{Cox.1994} with $S$=1/2\cite{Cox.1994}, $g=$2.0024\cite{Cox.1994}, $I$=1/2\cite{Cox.1994}, $A_{\parallel}$= -159.73 MHz\cite{Cox.1994}, and $A_{\perp}$ = -113.84 MHz\cite{Cox.1994} to calculate $\left< \sin^2\left(\theta/2\right) \right>_L$ by numerical evaluation of the pulse propagator, as described in the Supplemental Material\cite{Supinfo1.3}. The numerical approach is an extension to the analysis of the DEER intensity reported in Stepanov et al.~\cite{Stepanov.2016} as we are able to consider forbidden transitions and simultaneous driving of multiple transitions in our refinement. Examining the vanishing residual around the four major dips in Fig.~\ref{Fig1}d confirms \Ns{15} to mostly constitute the spin bath in sample A, an observation that is also valid for sample B with the higher nitrogen concentration (see Supplemental Material\cite{Supinfo1.5.2}). Using that for a \Ns{15} bath ($S=1/2$ and $g=2.0024$\cite{Cox.1994}) the factor $A \gamma_{\mathrm{NV}} \gamma$ in Eq.~(\ref{eq_contrast_1_compact}) becomes 0.292~MHz/ppm\cite{Stepanov.2016,Eichhorn.2019}, we extract the \Ns{15} concentrations given in Table \ref{tab:pishift_conc} from the quantitative fit with Eq.~(\ref{eq_contrast_1_compact}), as shown for sample A in Fig.~\ref{Fig1}d.

\begin{figure} [h!]
	\begin{center}
	\includegraphics[scale=1.0]{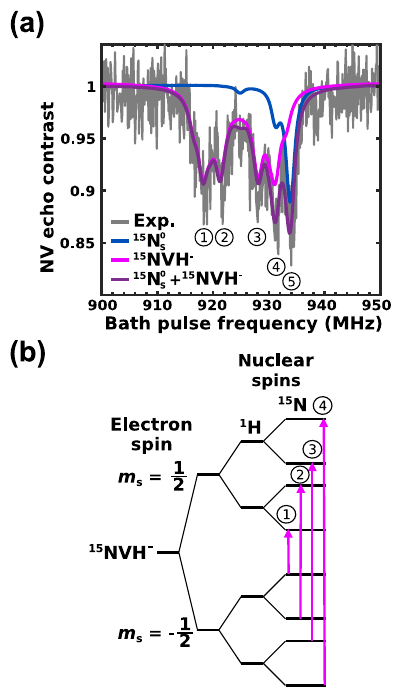}
	\end{center}
	\caption{a) High resolution DEER spectrum acquired at 330$\,$G aligned along $\left< 111 \right>$ and $\pi$-pulse length of 975$\,$ns using $\tau\,=\,$10$\,\mathrm{\mu}$s. The numbers 1$\,$-$\,$5 label the most prominent resonances in the spectrum and are assigned to the spin transitions represented by pink arrows in b). b) Level scheme of the $S = 1/2$ electron spin of \NVH{15}. The spin states $m_S = \pm 1/2$ get split each into  four sub-levels due to hyperfine interaction with $^1$H and $^{15}$N.}
	\label{Fig2}
	\end{figure}

By contrast, examining the spectrum in Fig.~\ref{Fig1}d between 900 and 950~MHz, we find a significant residual comparing model and experiment. Although considering a forbidden \Ns{15} transition around 930 MHz, the signal observed between 910$\,$MHz and 930$\,$MHz seems not be explained by \Ns{15} only. Interestingly, the same spectrum is also observed for Sample B, cf. Supplemental Material\cite{Supinfo1.5.2}. To further investigate the DEER intensity between 900$\,$MHz and 950$\,$MHz, we reduce the microwave power that the bath $\pi$-pulse is two-times longer than in Fig.~\ref{Fig1}d, i.e. 975$\,$ns and 525$\,$ns, respectively. By this, we address the bath spins more precisely with limited power broadening and find at least five resonances (1-5) (Fig.~\ref{Fig2}a), where peak no.~5 corresponds to the forbidden transition of \Ns{15} already evident in Fig.~\ref{Fig1}d. The remaining peaks, no.~1-4, seem to arise from another spin species and, therefore, we extend our model by adding the usually second most-abundant spin in $^{15}$N-doped diamond\cite{Edmonds.2012}, i.e. \NVH{15}, with the Hamiltonian $\hat{H}_{\mathrm{^{15}NVH}}=g\mu_B\vec{B}_{0}\hat{S}+\hat{S}\stackrel{\leftrightarrow}A_{\mathrm{^{15}N}}\hat{I}_{\mathrm{^{15}N}}+\hat{S}\stackrel{\leftrightarrow}{A_{\mathrm{^{1}H}}}\hat{I}_{\mathrm{^{1}H}}$ where $A_{\parallel,15N}$ = 2.9$\,$MHz\cite{Glover.2003}, $A_{\perp,15N}$ = 3.1$\,$MHz\cite{Glover.2003}, $A_{\parallel,1H}$= 13.69$\,$MHz\cite{Glover.2003}, $A_{\perp,1H}$ = -9.05$\,$MHz\cite{Glover.2003}. Assuming two randomly distributed spin species of concentrations $n_1$ and $n_2$, Eq.~\ref{eq_contrast_1_compact} extends then to:

\begin{equation}
C_{\mathrm{DEER}} =  C_{\tau} \exp\left(-A \gamma_{\mathrm{NV}} \gamma  T \left[  n_1x_1  + n_2 x_2 \right] \right) , 
\label{eq_contrast_two_species_compact}
\end{equation}

\noindent where $x_1$\,$=$\,$\left< \sin^2\left(\theta/2\right)\right>_{1,L}$ and $x_2 = \left< \sin^2\left(\theta/2\right)\right>_{2,L}$ are  the inverted fractions for spin species 1 and 2. We have made the approximation that both species have an isotropic and equal $g$ factor and therefore have the same gyromagnetic ratio $\gamma = g \mu_B/\hbar$.  Indeed, taking species 1 as \Ns{15} and species 2 as \NVH{15}, one obtains $|g_2/g_1 - 1|<|2.0034/2.0024-1| =  5.0\textrm{e}-4 \ll 1$, which validates the approximation. 

Including also \NVH{15} into the model, the fit agrees well with the experimental data in Fig.~\ref{Fig2}a and, therefore, we attribute the signal observed between 910$\,$MHz and 930$\,$MHz to \NVH{15} spins in our doped layer. The resonances no.~1-4 are well reproduced by the model, and thus can be assigned to the identically labelled electron spin transitions in Fig.~\ref{Fig2}b. Since the \NVH{15} center hosts two nuclear spins ($^{15}$N and $^1$H) each with $I$\,=\,1/2, the two electron spin states $m_S=\pm1/2$ split up each into four hyperfine states leading to four transitions, represented by pink arrows in Fig.~\ref{Fig2}b. The major resonances, 1-4 in Fig.~\ref{Fig2}a, correspond to the three equivalent orientations of \NVH{15} centers that are not aligned with $\vec{B}_{0}$. Since the effective hyperfine interaction between the electron and the nuclear spin ($^{15}$N or $^1$H) depends on the orientation of the defect with respect to the magnetic field, the transition frequencies differ for the aligned and the not-aligned \NVH{15} centers. In particular, the aligned subset experiences slightly larger $^1$H splittings, causing weak shoulders in Fig.~\ref{Fig2}a. Additional forbidden transitions contribute to the intensity in the central region around 925$\,$MHz but cannot be individually resolved. By repeating the measurements and the fitting procedure with Eq.~\ref{eq_contrast_two_species_compact} for sample B we observe the same hyperfine structure (Supplemental Material\cite{Supinfo1.5.2}) and Table~\ref{tab:pishift_conc} summarizes the obtained \NVH{15} concentrations for ensemble A1 and B1. The higher \NVH{15} concentration in sample B can be attributed to the overall higher nitrogen concentration in sample B than in A. As a consequence, the more dense spin bath in sample B explains also the broader Lorentzian linewidth for \NVH{15} spins, with (2.44\,$\pm$\,0.18)\,MHz and (1.33\,$\pm$\,0.11)$\,$MHz (FWHM) for ensemble B1 and A1, respectively. It should be noted that the linewidths given above are not affected by power broadening since the effect of the microwave power is considered in the model separately from intrinsic inhomogeneous broadening (Supplemental Material\cite{Supinfo1.3}). 

\begin{table}
\caption{\label{tab:pishift_conc}The concentrations of \Ns{15} and \NVH{15} obtained from fitting the DEER spectrum (spec.) and the $\pi$-shift method ($\pi$). For \Ns{15}, the allowed transitions as for example in Fig.~\ref{Fig1}d are fitted while for \NVH{15} only the region between 900$\,$MHz and 950$\,$MHz is used (Fig.~\ref{Fig2}a). The full set of fit parameters are provided in the Supplemental Material\cite{Supinfo1.5.2Tables}. In case of the $\pi$-shift method for \Ns{15} the experiments are performed on each resonance line (cf. Fig.~\ref{Fig1}d) and the results summed up. For \NVH{15}, the experimental data from a single $\pi$-shift experiment is fitted using Equation \ref{eq_contrast_two_species_compact} with the calculated fractions of inverted spins $x_{\mathrm{N}_s^0}$, $x_{\mathrm{NVH}^-}$ (Fig.~\ref{Fig3}a).}
\begin{ruledtabular}
\begin{tabular}{ccccc}
 Ensemble &$n_{\mathrm{N}_s^0}$ (ppm)& $n_{\mathrm{NVH}^-}$ (ppm)&$x_{\mathrm{N}_s^0}$& $x_{\mathrm{NVH}^-}$\\
A1 (spec.) & 7.66$\,$$\pm$$\,$0.12 & 0.40$\,$$\pm$$\,$0.02 & - & -\\
A1 ($\pi$) & 6.65$\,$$\pm$$\,$0.55 & 0.34$\,$$\pm$$\,$0.07 & 0.007 & 0.448\\
B1 (spec.) & 11.02$\,$$\pm$$\,$0.28 & 0.81$\,$$\pm$$\,$0.04 & - & -\\
B1 ($\pi$) & 9.90$\,$$\pm$$\,$0.87 & 0.67$\,$$\pm$$\,$0.14 & 0.004 & 0.597 \\
\end{tabular}
\end{ruledtabular}
\end{table}

So far, we have demonstrated that \NVH{15} can be detected in sub-micron thick $^{15}$N-doped diamond layers using optically detected DEER with NV$^{-}$ as atomic-sized sensors. The presented method, however, is also applicable to detect \NVH{14} in $^{14}$N-doped diamond. Studying a sample grown with identical parameters as sample A but using $^{14}$N as dopant, yields the DEER spectra in the Supplemental Material\cite{Supinfo1.5.1}. Due to the $I$~=~1 nuclear spin of $^{14}$N we observe an additional allowed transition arising from \Ns{14} that overlaps with half of the \NVH{14} spectrum\cite{Glover.2003}. Using the quantitative fitting method with Eq.~\ref{eq_contrast_two_species_compact} and the hyperfine tensor derived from \NVH{15}\cite{Glover.2003}, confirms successfully the measured DEER signal to arise from \NVH{14}.

\begin{figure*} [ht]
	\begin{center}
	\includegraphics[scale=1.0]{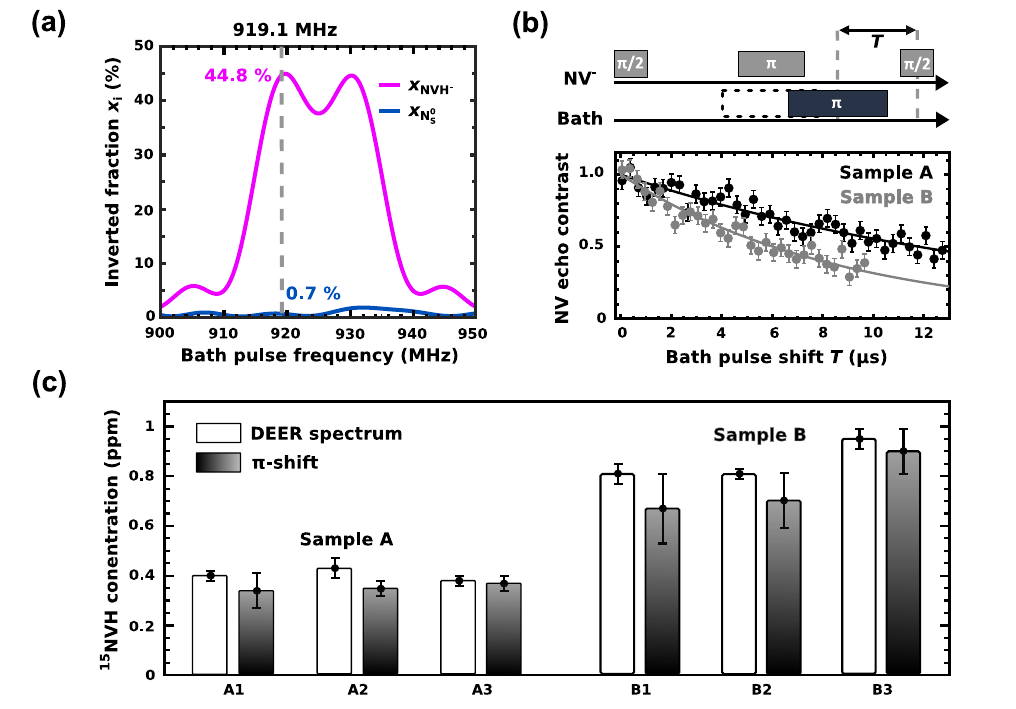}
	\end{center}
	\caption{a) The fraction of spins inverted by a 93.4$\,$ns $\pi$-pulse as a function of the pump frequency for ensemble A1. The calculation is performed with EasySpin using the linewidths and experimental parameters obtained from the fit in Fig.~\ref{Fig2}a. b) Pulse sequence of the $\pi$-shift method and experimental results for ensembles A1 and B1. The solid lines correspond to fits with mono-exponentials defined by Eq.~(\ref{eq_contrast_two_species_compact}). c) Comparison of the \NVH{15} concentrations obtained from the quantitative fit method for the high resolution spectra (Fig.~\ref{Fig2}a) and the $\pi$-shift method (see subfigure b) for six NV ensembles in Sample A and B. The uncertainties correspond to fit errors (95$\,\%$-confidence interval).}
	\label{Fig3}
	\end{figure*}

An alternative method for determining the spin bath concentrations is shifting the resonant bath $\pi$-pulse by $T$ with respect to one of the NV$^-$ $\pi/2$ pulses (Fig.~3b). The NV-echo contrast $C_{\mathrm{DEER}}$ follows then a mono-exponential decay given by Eq.~(\ref{eq_contrast_1_compact}) or Eq.~(\ref{eq_contrast_two_species_compact}), for the case of one or two bath spin species, respectively. For the allowed transitions of \Ns{15}, the spectral overlap with \NVH{15} is negligible (Fig.~\ref{Fig1}d), so that Eq.~(\ref{eq_contrast_1_compact}) can be used. This technique has been used in several studies to analyze the \Ns{} bath in diamond layers\cite{Eichhorn.2019,Osterkamp.2020,Balasubramanian.2022,Hughes.2023} and we here refer to it as $\pi$-shift method. Performing the $\pi$-shift method on each \Ns{} transition frequency and fitting Eq.~(\ref{eq_contrast_1_compact}) to the experimental data in the Supplemental Material\cite{Supinfo1.5.2FigS10}, we extract a total \Ns{15} bath concentration of (6.7$\,$$\pm$$\,$0.6)$\,$ppm and (9.9$\,$$\pm$$\,$0.9)$\,$ppm, for sample A and B, respectively (Table S9). The \Ns{} concentrations obtained from the quantitative fits of the spectra in Table~\ref{tab:pishift_conc} agree well with the results from the $\pi$-shift method. 

As a result, one should be able to estimate the \NVH{15} concentration in the same way. However, unlike for \Ns{}, the \NVH{15} resonances lie spectrally very close together and overlap with the forbidden transitions of \Ns{15} (Fig.~\ref{Fig2}a). Additionally, the expected concentration of NVH$^-$ is smaller by one order of magnitude than that of \Ns{15} (Table~\ref{tab:pishift_conc}). Hence, we apply a $\pi$-pulse with a duration of 93$\,$ns and frequency of 919.1$\,$MHz that we drive simultaneously transition no. 1 and 2 in Fig.~\ref{Fig2}a to enhance the DEER contrast from \NVH{15} spins. Fig.~\ref{Fig3}a shows the calculated inverted fractions $x_{\mathrm{N}_s^0}$ and $x_{\mathrm{NVH}^-}$ in Eq.~\ref{eq_contrast_two_species_compact} applying a 93$\,$ns long $\pi$-pulse depending on the bath pump frequency. For the calculation of $x_i$ we use in EasySpin the linewidths obtained from the fit in Fig.~\ref{Fig2}a for ensemble A1. Due to the strong $\pi$-pulse in Fig.~\ref{Fig3}a, transition no.$\,$1 and 2 merge to one peak and applying the $\pi$-pulse with 919.1$\,$MHz results in a spin inversion of 44.8$\,\%$ and 0.7$\,\%$ for \NVH{15} and \Ns{15}, respectively. Varying the pulse timing $T$, the NV echo contrast in Fig.~\ref{Fig3}b follows a mono-exponential decay that is fitted with Eq.~(\ref{eq_contrast_two_species_compact}). For the \Ns{15} concentration we use the value from the $\pi$-shift method (Table~\ref{tab:pishift_conc}) and obtain for \NVH{15} (0.34$\,\pm\,$0.07)$\,$ppm. Comparing with the concentration, (0.40$\,\pm\,$0.02)$\,$ppm, previously obtained from the quantitative fit of the spectrum in Fig.~\ref{Fig2}a, it can be seen that both methods agree with each other. Our results demonstrate that the $\pi$-shift method is applicable also in the case of overlapping resonances of two species, provided the concentration of one of them has been separately determined. Repeating the analysis for sample B with the higher nitrogen content, we observe a faster decay than for sample A in Fig.~\ref{Fig3}b resulting after analysis in a significant higher \NVH{15} concentration of (0.67$\,\pm\,$0.14) ppm. As for sample A, the result agrees well with the (0.81$\,$$\pm$$\,$0.04)$\,$ppm from the quantitative fit of the DEER spectrum. Repeating the spin bath analysis for the NV ensembles A2,3 and B2,3 yields within one method consistent \NVH{15} concentrations for sample A and B. Comparing the results of both approaches, however, we find systematically higher values for the DEER spectrum than for the $\pi$-shift method (Fig.~\ref{Fig3}c). 

The main difference between the spectrum and the $\pi$-shift method is the applied microwave power to excite the spin bath. In case of ensemble A1, for instance, using a ten-times shorter $\pi$-pulse for acquiring the signal in Fig.~\ref{Fig3}b (93$\,$ns) than for the spectrum in Fig.~\ref{Fig2}a (975$\,$ns), the contribution to the DEER intensity from off-resonant driving of allowed \Ns{15} transitions is expected to be bigger for the $\pi$-shift method than for the DEER spectrum. In Fig.~\ref{Fig3}a we consider the non-resonant excitation of \Ns{15} spins, but we assume rectangular-like $\pi$-pulses for calculating $x_i = \left< \sin^2\left(\theta_i/2\right) \right>_{L}$ in EasySpin. As in the experiment the pulse shape can deviate from our assumption, the calculated inverted spin fractions might be not exact for high microwave powers leading to a possible overestimation of the contribution from \Ns{15} to the DEER intensity in Fig.~\ref{Fig3}b. But considering the small difference between the concentrations from the spectrum and the $\pi$-shift method in Fig.~\ref{Fig3}c, the assumption of rectangular-like pulses seems to be a reasonable approximation for determining the \NVH{15} concentration with a short, high power microwave pulse in Fig.~\ref{Fig3}a+b. In fact, a higher microwave power results in a stronger NV echo contrast and, therefore, shorter data acquisition time. The DEER spectrum for sample A in Fig.~\ref{Fig2}a, for example, has been measured for more than one day while the $\pi$-shift method like in Fig.~\ref{Fig3}b takes only a few hours. Once the typical linewidth of the NVH$^-$ spins in a given sample is determined, the inverted fractions can be calculated, and one could rely on the $\pi$-shift method only to analyze the spin bath around various NV$^-$ ensembles in an time-efficient way. This works under the assumption that large variations of the spin concentration can be excluded as this might influence the linewidths. For heterogeneous diamond layers, one should stick to the quantitative fit of the DEER spectrum as it considers directly the linewidths of the spin species and is therefore more robust.

For samples with defect concentrations lower than in the present study, the pulse spacing $\tau$ in Fig.~\ref{Fig1}c has to be increased for both approaches. To maximize the signal measured in a given time, it should be chosen such that $2\tau \approx T_{2,\mathrm{NV}}$, as this ensures a high DEER intensity ($T \approx \tau$ in Eq.~\ref{eq_contrast_two_species_compact}) while preserving enough NV echo contrast and thus a sufficient signal-to-noise ratio. Since we expect typically a longer $T_{2,\mathrm{NV}}$ in diamonds with lower nitrogen content, matching the $2\tau \approx T_{2,\mathrm{NV}}$ condition will result in a sensitivity improvement for DEER partially compensating the overall lower density of bath spins to be detected. Additionally, dynamical decoupling sequences\cite{Alvarez.2010} could enhance $T_{2,\mathrm{NV}}$ even further and, as a consequence, the DEER techniques presented in this work are not restricted to the nitrogen concentrations of sample A and B, 13 and 17 ppm respectively, and can be therefore applied to even more dilute N-doped diamond layers of in principle any thickness.

In conclusion, we have shown that optically detected DEER with NV$^-$ centers can be applied for detecting and investigating NVH$^-$ centers in sub-micron-thick diamond layers. Since the method relies on  NV$^-$ centers as local magnetic field sensors, optically detected DEER is decoupled from the sample volume and works, therefore, even on the nanoscale. The ability to estimate the concentration of NVH$^-$ in the range of a few hundreds of ppb, opens the possibility to compare and engineer spin bath environments aiming for higher N-to-NV conversion ratios without the need for thick bulk samples. This might boost the understanding of the defect generation during CVD diamond growth and pave the way for better diamond material for quantum applications. Furthermore, being sensitive not only to \Ns{} but also \NVH{}, optically detected DEER could be employed for imaging dark spins using a wide-field microscope offering entirely new insights into defect correlation and material heterogeneity on the mesoscale.

\section{Acknowledgements}
The authors thank Philipp Vetter for experimental support and acknowledge also the support by Cambridge Isotope Laboratories Inc.. This work was funded by "European Research Council" (HyperQ GAno. 856432), "Bundesministerium für Bildung und Forschung" Germany (QSCALE Fkz. 03ZU1110GB, QSPACE Fkz. 03ZU1110JA), and "Land Baden-Württemberg" (QC-4BW Az. 3-4332.62-IAF/7, QC-4BW-2 Az. WM3-4332-96/10).\\
\bibliography{references}

\pagebreak


\setcounter{figure}{0}

\setcounter{table}{0}
\renewcommand{\thetable}{S\arabic{table}}
\setcounter{figure}{0}
\renewcommand{\thefigure}{S\arabic{figure}}
\setcounter{equation}{0}
\renewcommand{\theequation}{S\arabic{equation}}






\onecolumngrid

\section*{Supplemental Material} 

\subsection{Chemical vapor deposition of diamond}
The samples in this work are grown using a home-built CVD system\cite{Silva.2010}. The parameters and thicknesses are summarized in Table \ref{tab:CVDparam}. For Sample C, a layer thickness and nitrogen concentration comparable to Sample A is expected, as the CVD process was the same.

\begin{table}[h!]
\caption{\label{tab:CVDparam}CVD nitrogen concentrations and the resulting thicknesses from SIMS (Figure S\ref{figS1:SIMS}). The chamber pressure was $p$=72.5\,mbar, the microwave power P=1.2\,kW, the sample holder temperature 785\,°C, the hydrogen and methane flow 600 and 3\,sccm, respectively.\\}
\begin{tabular}{cccccc}
\hline
\hline
 Sample & Isotope & N:C ratio & $[N_{\mathrm{gas}}]$ (ppm)& $d_{\mathrm{layer}}$ (nm) & $[N_{\mathrm{diamond}}] $ (ppm) \\
\hline
A & $^{15}$N & 1.25 & 3100 & 570$\,\pm\,$60 & 13.8$\,\pm\,$1.6\\
B & $^{15}$N & 2.0 & 5000 & 700$\,\pm\,$50 & 16.7$\,\pm\,$3.6\\
C & $^{14}$N & 1.25 & 3100 & n.a. & n.a. \\
\hline
\hline
\end{tabular}
\end{table}

\begin{figure}[H]
	\centerline{\includegraphics[scale=0.4]{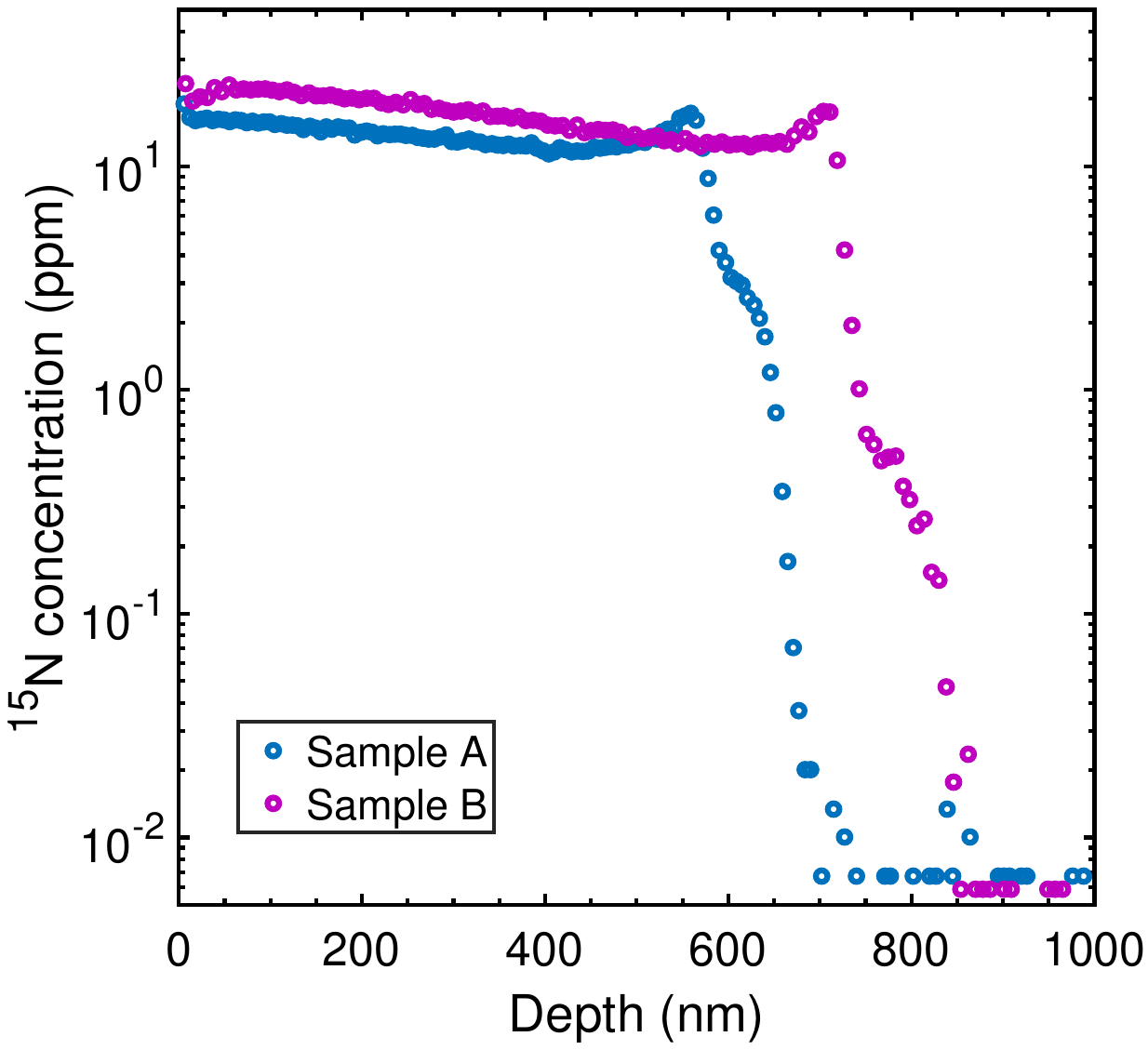}}
	\caption{$^{15}$N concentration profiles obtained from secondary ion mass spectrometry (SIMS) for Sample A and B. The average concentration for Sample A and B is (13.8$\,\pm\,$1.6)\,ppm and (16.7$\,\pm\,$3.6)\,ppm, respectively, where the uncertainty comes from the width of the transition region marked by the steep increase in $^{15}$N content going over from the substrate (background level 10$^-2$\,ppm) and to the CVD layer.}
	\label{figS1:SIMS}
\end{figure}

\subsection{Expression for the DEER intensity}

Although the photoluminescence-based readout method used in the present work differs from that in conventional EPR experiments, using inductive detection, the underlying physics is the same and the same equations  apply to describe the DEER signal.   
We mostly use the findings by  
Stepanov and Takahashi~\cite{Stepanov.2016}, that detailed the theory of the DEER experiment in conditions of  high-frequency/high-field EPR (115\,GHz / 4.1\,T). 
We adapt in particular the formula made to describe the DEER signal, to the case of the present experiment. From that earlier work, the DEER intensity reads:

\begin{equation}
C_{\mathrm{DEER}} = C_0  \exp\left(-\frac{2\pi\mu_0\mu_B^2g_{\mathrm{NV}} g  T}{9\sqrt{3}\hbar}n_{\mathrm{bath}} \left< \sin^2\frac{\theta}{2} \right>_L\right) \exp\left( -\frac{2\tau}{T_2} \right),
\label{eq_contrast_1_species}
\end{equation}

\noindent where  $\theta$ is the flip angle induced by the pulse on the bath spins, $n_{\mathrm{bath}}$ is the spin concentration,   $\left< \right>_L$ means averaging is performed over the full, inhomogeneously broadened, spectrum (so that $\left<\sin^2\frac{\theta}{2} \right>_L$ quantifies the overall efficiency of the inversion pulse on the bath). $g_{\mathrm{NV}}, g$ are the $g$ factors of the NV and the bath spins, respectively, the term $\exp\left( -\frac{2\tau}{T_2} \right)$ accounts for the loss of coherence of the NV center during the sequence  that is not related to the partial inversion of the bath spins, $C_0$ is the maximal contrast obtained on the NV centers. We note that $C_0$ includes the effects of unperfect driving caused, e.g., by inhomogeneous broadening of the NV center line.  \\

\textbf{Adding one bath spin species}. We now consider the case when the local environment of the NV center consists of two other species, with concentration $n_1$ and $n_2$ respectively, that are randomly distributed. Equation (\ref{eq_contrast_1_species}) becomes:

\begin{equation}
\scalebox{0.8}{\mbox{\ensuremath{\displaystyle 
C_{\mathrm{DEER}} = C_0 \exp\left(-\frac{2\pi\mu_0\mu_B^2g_{\mathrm{NV}}  T}{9\sqrt{3}\hbar} \left[  n_1 g_1 \left< \sin^2\frac{\theta}{2} \right>_{1,L}  + n_2 g_2 \left< \sin^2\frac{\theta}{2} \right>_{2,L} \right] \right) \exp\left( -\frac{2\tau}{T_2} \right), }}} 
\label{eq_contrast_2_species}
\end{equation}

\noindent which reflects the fact that, with the DEER sequence,  the pulse on the bath spins for species $i$ yields a contribution to the decoherence rate of the NV center, that scales with the value $\left<\sin^2\frac{\theta}{2} \right>_{i,L}$, the concentration $n_i$, and the $g$ factor $g_i$. Isolating the terms that can vary in our experiments, we rewrite the equation into the more compact formula:

\begin{equation}
C_{\mathrm{DEER}} =  C_{\tau} \exp\left(-A \gamma_{\mathrm{NV}} \gamma  T \left[  n_1 x_1  + n_2 x_2 \right] \right) , 
\label{eq_contrast_2_species_compact1}
\end{equation}

\noindent where $ x_1 = \left< \sin^2\frac{\theta}{2} \right>_{1,L}$, $ x_2 = \left< \sin^2\frac{\theta}{2} \right>_{2,L}$,  $C_{\tau} - C_0\exp\left( -\frac{2\tau}{T_2}\right)$, $A=\frac{2\pi\mu_0 \hbar  }{9\sqrt{3}}$.  $\gamma_{\mathrm{NV}}=\frac{g_{\mathrm{NV}} \mu_B}{\hbar}$,  $ \gamma=\frac{g \mu_B}{\hbar}$ are the gyromagnetic ratios for the NV and for the bath spins respectively. We made the approximation of identical $g$-factors for the two species  ($g_1 = g_2 = g$) yielding identical gyromagnetic ratios ($\gamma_1 = \gamma_2 = \gamma$). Taking species 1 as P1 and species 2 as NVH$^-$, despite the $g$-factor anisotropy for NVH$^-$ ($g_{\parallel}=2.0034$, $g_{\perp}=2.0030$)\cite{Glover.2003}, one has (Table~\ref{table_Hamiltonian_param_values})
$| \frac{g_2}{g_1} - 1|<|\frac{2.0034}{2.0024}-1| =  5.0\textrm{e}-4$, which validates the approximation.  \\

\subsection{Calculation}

The main task in the derivation of the DEER intensity, from  Eq.~(\ref{eq_contrast_2_species_compact1}) is,  calculating the term $\left< \sin^2\frac{\theta}{2} \right>_{i,L}$ for each spin species $i$.  We here present analytical expressions found in the literature, discuss limitations to their validity range, and how a refined description combined with a numerical approach allow for a more robust and accurate treatment of the problem.

\subsubsection{Analytical expressions}

The expression for the $\left< \sin^2\frac{\theta}{2} \right>_{i,L}$ term in the case of an inhomogeneously broadened spectrum (described by the shape $L(\omega)$) has been derived in ref.~\cite{Salikhov.1981} as: 

\begin{equation}
\left< \sin^2\frac{\theta}{2} \right>_{L} = \int_{-\infty}^{\infty} \frac{\Omega_1^2}{\Omega_1^2+
(\omega-\omega_0)^2}  \sin^2\left(  \frac{t_p}{2} \sqrt{\Omega_1^2 + (\omega-\omega_0)^2}) L(\omega)\right) d\omega,
\label{eq_sin2_salikhov1981}
\end{equation}

\noindent where  $t_p$ and $\omega_0$ are the duration and carrier frequency of the microwave pulse, respectively, $\Omega_1$ is the on-resonance nutation frequency of the electron spins. This expression assumes all spin transitions have the same $\Omega_1$. We now generalize it to the case when $\Omega_1$ varies across the spectrum, which, importantly, allows describing the contribution from both allowed and  possible forbidden transitions:

\begin{equation}
\left< \sin^2\frac{\theta}{2} \right>_{L} = \Sigma_{\mathrm{a,b}} \int_{-\infty}^{\infty} \frac{\Omega_{1,\mathrm{ab}}^2}{\Omega_{1,\mathrm{ab}}^2+
(\omega-\omega_0)^2}  \sin^2\left( \frac{t_p}{2}   \sqrt{ \Omega_{1,\mathrm{ab}}^2 + (\omega-\omega_0)^2}\right) L_{\mathrm{ab}}(\omega) d\omega,
\label{eq_sin2_weakdrive_var_omega1}
\end{equation}

\noindent where the sum runs over all possible transitions, $\mathrm{a} \leftrightarrow \mathrm{b}$,  between the $S_z = -1/2$ and $S_z = 1/2$ subspaces of the defect, $L_{\mathrm{ab}}(\omega)$ describes both the  position in the spectrum and the homogeneous broadening corresponding to the $\mathrm{a} \leftrightarrow \mathrm{b}$ transition (so that doing the summation, $L(\omega) = \Sigma_{\mathrm{a,b}} L_{\mathrm{ab}}(\omega)$, allows to recover the complete lineshape), $\Omega_{1,\mathrm{ab}}$ is the transition-specific on-resonance Rabi frequency. For a linearly polarized field $\vec{B_1}$ (of arbitrary direction) acting on a $S=1/2$ electron, it reads  $\Omega_{1,\mathrm{ab}} = \gamma |\bra{b} \vec{B_1}\cdot \vec{S} \ket{a}|/2$, where $\gamma$ is the gyromagnetic ratio of the considered spin.

\subsubsection{Effect of driving strength}

\begin{figure}[htp]
	\centerline{\includegraphics[width=300pt]{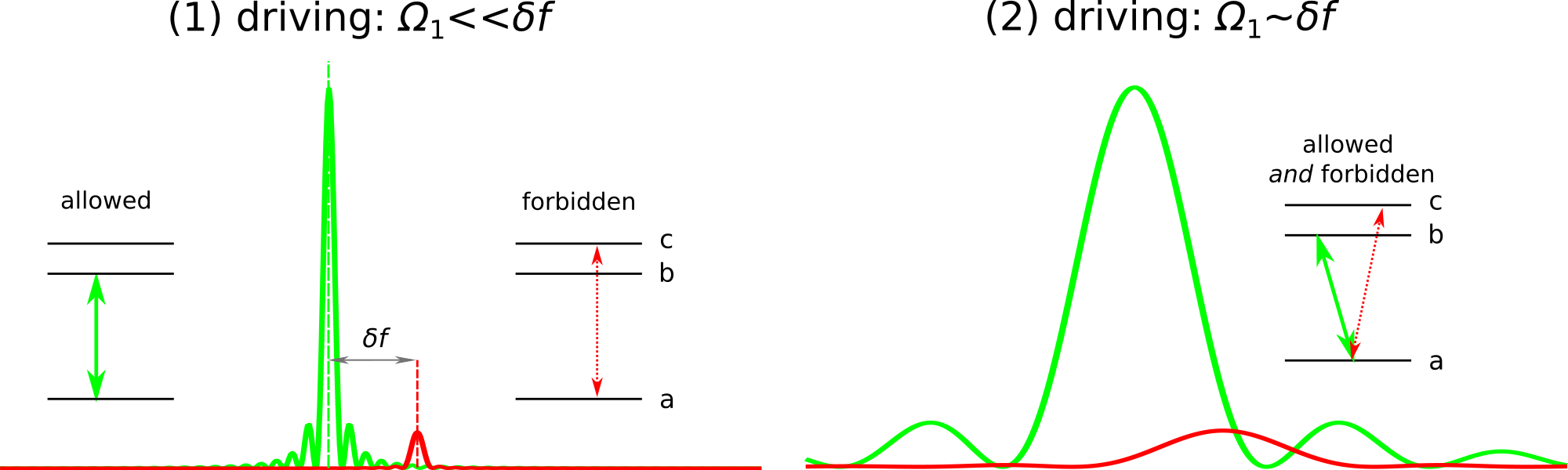}}
	\caption{Description of the two driving regimes in the case of NVH$^-$, or for any other center presenting both an allowed and a forbidden spin transition ($a \leftrightarrow b$ and $a \leftrightarrow c$, respectively) detuned by $\delta f$. The contribution of each transition to the sum in Eq.~(\ref{eq_sin2_weakdrive_var_omega1}) is plotted (inhomogeneous broadening is ignored). Significant overlap of the two curves in the case (2) $\Omega_1 \sim \delta f$ signals in fact the limit of validity of this description, justifying the label of `weak driving approximation'. }
	\label{fig_sin2_allowed_forbidden_NVH}
\end{figure}

\begin{figure}[h!]
	\centerline{\includegraphics[width=400pt]{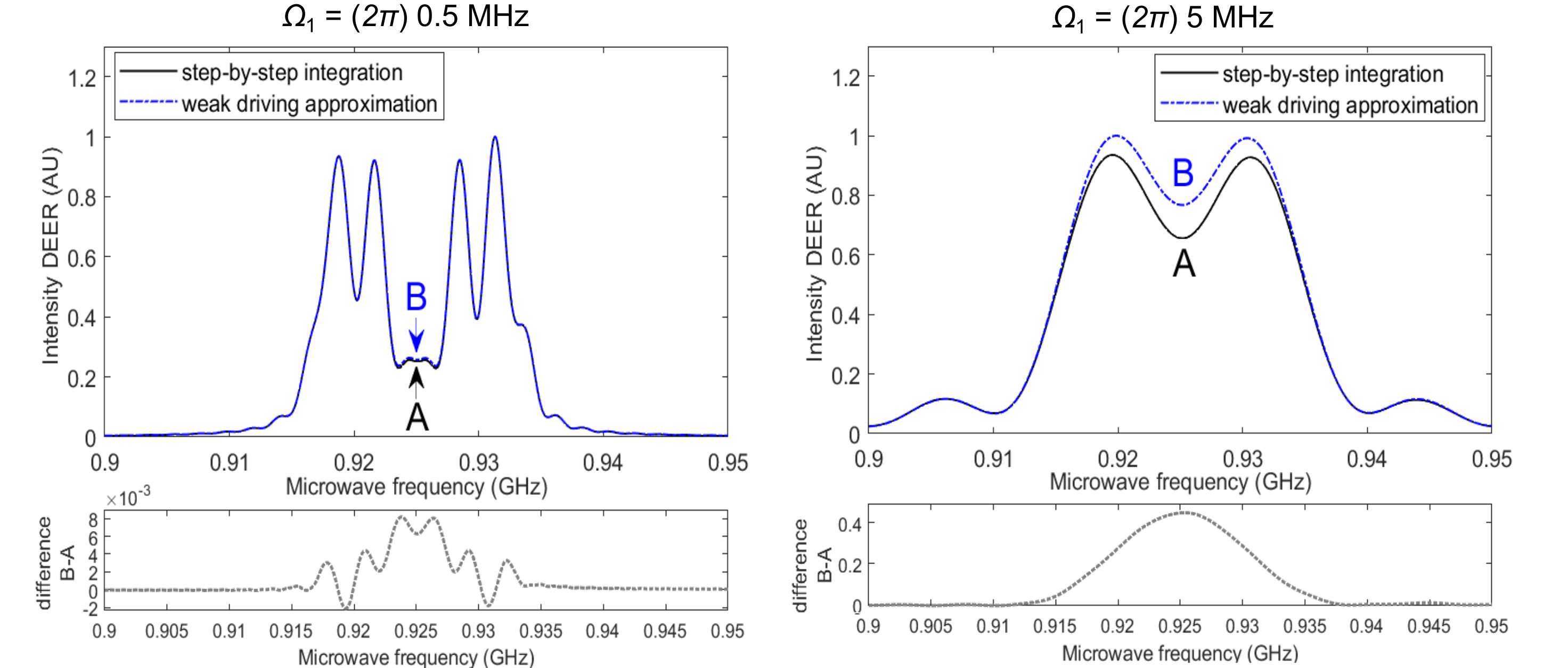}}
	\caption{Numerical simulations of the DEER intensity for NVH$^-$ at $B=33$\,mT in the cases $\Omega_1 = (2\pi) 0.5\,$MHz and $\Omega_1 = (2\pi) 5\,$MHz, using either the step-by-step integration approach or the weak driving approximation (A, B on each plot). }
	\label{fig_weak_vs_strong_driving_2figs}
\end{figure}

\begin{table}[h!]
\centerline{
\begin{tabular}{cccc}
\hline 
& & P1($^{15}$N)~\cite{Smith.1959,Cox.1994} & $^{15}$NVH$^-$\cite{Glover.2003}  \\ \hline 

$g$-factor			      &                          & $g = 2.0024$      &     $g_{\parallel}=2.0034,g_{\perp}=2.0030$        \\  \hline

$^{15}$N hyperfine        & $A_{\parallel}$ (MHz) & -159.73   &  2.94       \\ 
                		  & $A_{\perp}$ (MHz)     & -113.84   &  3.10       \\   \hline
$^{1}$H hyperfine         & $A_{\parallel}$ (MHz) & -  		  &  13.69       \\ 
                		  & $A_{\perp}$ (MHz)     & -  		  &  -9.05       \\   \hline    	 

\end{tabular}}
\caption{Hamiltonian parameters for the spin species and references. }
\label{table_Hamiltonian_param_values}
\end{table}

\paragraph{Weak vs strong driving.} We find Eq.~(\ref{eq_sin2_weakdrive_var_omega1}) to be valid only up to a certain strength of $B_1$ field, as we will now describe. We first point out that, doing the summation over all transitions in Eq.~(\ref{eq_sin2_weakdrive_var_omega1}), implicity assumes that two distinct transitions are always driven independently. In other words, it is assumed that the situation should not occur,  when an allowed transition $\mathrm{a} \leftrightarrow \mathrm{b}$ is driven while $\mathbf{a \leftrightarrow c}$, a forbidden one, is also manipulated by the microwave excitation. This assumption is not true, in the case of NVH$^-$, for which the allowed and forbidden transitions are weakly separated in frequency (by $\delta f = 5-10\,$MHz at $\sim 30$\,mT magnetic field), so that they can be simultaneously affected by the drive of a single pulse in our experiment. As $\Omega_{1,\mathrm{ab}}>\Omega_{1,\mathrm{ac}}$, the possibility of simultaneous driving exists for $\Omega_{1,\mathrm{ab}} \sim \delta f$ or at higher values of $\Omega_{1,\mathrm{ab}}$. The occurence of the treshold, at which the formula in Eq.~(\ref{eq_sin2_weakdrive_var_omega1}) becomes invalid due to simultaneous driving,  is depicted in Fig.~\ref{fig_sin2_allowed_forbidden_NVH}. In the figure, the individual terms being summed in the equation are represented, for varying $\Omega_{1,\mathrm{ab}}$, always assuming $\Omega_{1,\mathrm{ac}} = 0.2 \, \Omega_{1,\mathrm{ab}}$ (that is taking the dipole moment $\sim 5$ times lower for the forbidden transition, as can be the case for NVH$^-$). Overlap of the corresponding curves in the spectral region in between the two transitions reveals that the approach taken  in  Eq.~(\ref{eq_sin2_weakdrive_var_omega1}) is in fact invalid.  Therefore, we will refer to performing the summation in Eq.~(\ref{eq_sin2_weakdrive_var_omega1}) as following the ``weak driving approximation''.

\paragraph{Strong driving case, numerical integration.}  In the case of stronger driving fields $\Omega_{1,\mathrm{ab}} \gtrsim \delta f$, an alternative way of computing the DEER signal is required, that we now present. We consider the density matrix of a bath spin species in its thermal state, that is $(\rho_{jk})$, with $\rho_{jk}=0$ for $j \neq k$ and $\rho_{jj}= \exp \left( -\frac{E_j}{k_B T} \right) /\left( \Sigma_{j'} \exp \left( -\frac{E_{j'}}{k_B T} \right) \right)$, $E_j$ is the energy of the spin eigenstate $j$, $k_B$ the Boltzmann constant and $T$ here the temperature. We now write:  

\begin{equation} 
\displaystyle
\left< \sin^2\frac{\theta}{2} \right>_L =  \int_{-\infty}^{+\infty}  \left(  \sum_j  \rho_{jj}  \left( \sin^2\frac{\theta}{2} \right)_{\mathrm{state}\, j, \omega'}  \right) f_L(\omega-\omega') d \omega',
\label{eq_sin2_density_matrix}
\end{equation}

\noindent where $\left( \sin^2\frac{\theta}{2} \right)_{\mathrm{state}\, j, \omega'}$ accounts for the case when the bath spins are initially found in the eigenstate $j$, and the pulse on the bath spins is sent at the frequency $\omega'$, $f_L$ is a line broadening function.

For illustration, we first consider the simplest case (although it can also be solved using the ``weak driving approach'') of a free electron with the Hilbert space consisting of the states defined by the two possible eigenvalues of $S_z$: $\ket{-\frac{1}{2}}$, $\ket{\frac{1}{2}}$. Assuming a spin is to be found initially in state $\ket{-\frac{1}{2}}$, application of a resonant microwave pulse makes it evolve into a state of the form  $\cos(\frac{\theta}{2})\ket{-\frac{1}{2}} + \sin(\frac{\theta}{2}) e^{i\varphi}\ket{\frac{1}{2}}$. In the symmetrical case, starting in the state
$\ket{\frac{1}{2}}$, evolution leads the system into the state $-\sin(\frac{\theta}{2}) e^{i\varphi} \ket{-\frac{1}{2}} + \cos(\frac{\theta}{2}) \ket{\frac{1}{2}}$ (the interaction picture is assumed in both cases). We remark that these two scenarios will always occur in a given spin bath, with a probability given by the Boltzmann weights ($\rho_{jj}$). In terms of density matrices, the first case, for instance,  corresponds to the evolution:

\begin{equation}
  \begin{pmatrix}
    1 & 0  \\
    0 & 0 
  \end{pmatrix}
    \rightarrow 
  \begin{pmatrix}
    \cos^2(\frac{\theta}{2})                                     & \sin(\theta) e^{-i\varphi} /2                    	\\
    \sin(\theta) e^{i\varphi}/2     & \sin^2(\frac{\theta}{2})
  \end{pmatrix},
\label{eq_density_matrix_simple_evolution}
\end{equation}
\noindent where we used the simplification  $\sin(\frac{\theta}{2})  \cos(\frac{\theta}{2}) =  \frac{1}{2} \sin(\theta) $. Thus, evolution of the density matrix yields the term to be summed in  Eq.~(\ref{eq_sin2_density_matrix}). To get an expression that is valid for both initial states $\ket{-\frac{1}{2}}$, $\ket{\frac{1}{2}}$, it is convenient to use the expectation value of the $S_z$  spin operator. If $m^{j,-}$ is the density matrix for the spin initially in state $j$ (left hand term in Eq.~(\ref{eq_density_matrix_simple_evolution})), and $m^{j,+}$ is the evolved density matrix (right hand term in Eq.~(\ref{eq_density_matrix_simple_evolution})), one can write: 
\begin{equation}
 \sin^2\left(\frac{\theta}{2}\right) = |\mathrm{tr}(m^{j,+}S_z) - \mathrm{tr}(m^{j,-}S_z)|. 
\label{eq_sin2_Sz_corresp}
\end{equation}

This expression is valid assuming both $\ket{-\frac{1}{2}}$, $\ket{\frac{1}{2}}$ cases of the initial spin state.  Furthermore, since the DEER intensity is provided by the action of a microwave pulse on the electron spin, rather than on any nuclear spin, this approach can be generalized to more complex spin species where (hyperfine-coupled) nuclear spins are also present, such as, in P1 and NVH$^-$.  Importantly, it remains that in Eq.~(\ref{eq_sin2_Sz_corresp}), one considers  the evolution of partial density matrices  $m^{j,-}$  (corresponding to pure initial states $j$), rather than the evolution of the \textit{full} density matrix. Intrinsically, the reason is, that DEER signal occurs as soon as there is change in the statistical spin bath polarization, following application of a microwave pulse, regardless of the initial average polarization. Considering the full density matrix would yield only the  average quantities, instead of quantities that are  dependent on statistical effects, such as the DEER intensity.  
The density matrix $m^{j,+}$ can be obtained as  $m^{j,+} = U m^{j,-} U^{\dagger}$ where U is the pulse evolution operator.

Although Eq.~(\ref{eq_sin2_Sz_corresp}) is valid in most cases for the spin species P1, NVH$^-$, it needs to be adapted to take into account a particular effect at the field of operation that is considered in the present work. When the strength of the hyperfine coupling is close enough to the electron Zeeman energy  
  (this is the case here for P1, see the value of the hyperfine interaction terms in Table~\ref{table_Hamiltonian_param_values}) 
mixing of the electron spin states can occur, which results in non-zero transition dipoles for a field oriented along $z$. As a consequence, the expectation value of the $S_z$ operator following the application of a microwave pulse, $\mathrm{tr}(m^{j,+}S_z)$, can be time-dependent when a forbidden transition is excited, with an oscillatory behavior  (same as, for an allowed transition, expectations values of $S_x$, $S_y$  when the inversion is not full). This oscillatory behavior follows the frequency $\nu_{\alpha\beta}$ of the forbidden transition as it reflects coherence between the two states involved. This coherence will however not lead to any effect on the DEER intensity. The latter  results from phase accumulation on the probe spin over a duration $\sim\tau$, and thus requires, in fact, using  a time-averaged quantity such as  $\braket{\mathrm{tr}(m^{j,+}S_z)}_{\tau}$. As $\nu_{\alpha\beta} \tau \gg 1$, the contribution from the oscillating, coherent, term vanishes. We therefore calculate the time-averaged quantity  by removing the coherent (that is, off-diagonal) terms in the matrix $m^{j,+}$:

\begin{equation}
\braket{\mathrm{tr}(m^{j,+}S_z)}_{\tau} =  \mathrm{tr}(m^{j,+,\mathrm{diag}}S_z),
\label{eq_tau_average}
\end{equation}

\noindent where $m^{j,+,\mathrm{diag}}$ is the matrix obtained by removing all off-diagonal  elements from 
the matrix $m^{j,+}$. One can finally write using this purely diagonal matrix:

\begin{equation}
 \sin^2\left(\frac{\theta}{2}\right) = |\mathrm{tr}(m^{j,+,\mathrm{diag}}S_z) - \mathrm{tr}(m^{j,-}S_z)|. 
\label{eq_sin2_Sz_diag}
\end{equation}

Practically the computation is performed using functions from the Matlab package Easyspin~\cite{Stoll.2006}. 
  For each species, the static Hamiltonian is computed using the \texttt{sham} function, while the driving field Hamiltonian is simply expressed as a function of the electron spin operator: $H_1 = \gamma  \vec{B_1}(t) \cdot \vec{S}$, where $\vec{B_1}(t) = B_1 \cos(\omega t) \vec{v}$, $\vec{v}$ is the direction of the linearly polarized field.  The pulse evolution operator $U$ is computed using the \texttt{propint} function, which performs step-by-step integration of the Liouville equation in the presence of the static and the time-dependent (cosine-modulated) Hamiltonian.  The combination of the \texttt{sham} and \texttt{propint} functions allows calculating each $m^{j,+,\mathrm{diag}}$. The matrices are then fed in a calculation using equation (\ref{eq_sin2_Sz_diag}), to provide an estimate for $\sin^2\left(\frac{\theta}{2}\right)$.
  
  In useful cases, the inhomogenous broadening $\left< \right>_L$ must be taken into account. To do so, we perfom the calculation of $\sin^2\left(\frac{\theta}{2}\right)$ at different microwave frequency detunings and performed weighted average with  the specified lineshape $L(\omega)$, to obtain  $\left< \sin^2\frac{\theta}{2} \right>_L$. For simulating a spectrum, rather than reproducing this weighted averaging step at each microwave frequency, it is more efficient to calculate  $\sin^2\left(\frac{\theta}{2}\right)$ for the full frequency range and perform then convolution with the lineshape function $L(\omega)$.  From there, one can use Eq. (\ref{eq_contrast_1_species}) (or, in the case of  two bath species, Eq. (\ref{eq_contrast_2_species_compact1})) to simulate the DEER intensity. 
  
  This approach, that we label step-by-step integration, is compared to the output of the ``weak driving approximation'' in Fig.~\ref{fig_weak_vs_strong_driving_2figs} for the case of NVH$^-$. It shows, as expected, that the approximation fails at high Rabi drive ($\Omega_1 = (2\pi)5$\,MHz on Fig.~\ref{fig_weak_vs_strong_driving_2figs}). In practice, we find that at  $\Omega_1 > (2\pi) 1$\,MHz, the other approach, although a little slower in terms of computation, should be preferred to compute NVH$^-$-related quantities. For P1, higher Rabi drives can be considered and the approximation approach still be valid, due to the more important spectral separation between the allowed and forbidden transitions.

\subsubsection{Fits}

Least-square fittings were performed using a Nelder/Mead downhill simplex algorithm (\texttt{esfit\_simplex} function from Easyspin) with a tolerance set to $1\mathrm{e}-3$. 

We include the static magnetic field $B_0$ (Zeemann field) and three angles into the fit functions. 
The first angle $\theta_{\mathrm{MW}}$ represents  the angle between the magnetic field $B_1$ used for microwave excitation and the static magnetic field $B_0$. The two other angles ($\delta_0$, $\phi_0$) correspond respectively to the polar and azimuthal angles giving the orientation of the external magnetic field $\vec{B_0}$ in spherical coordinates, in the frame related to the diamond crystal axes. A magnetic field aligned to the $\langle111\rangle$ crystal orientation, corresponds to  $\delta_0=54.74$°, $\phi_0=45$°. 
We note that to fully define the experiment geometry, a fourth angle could be used for defining the microwave field orientation: indeed, with  $\theta_{\mathrm{MW}}$ as a  polar angle, the definition of an azimuthal angle ($\phi_{\mathrm{MW}}$) would set exactly the orientation of $\vec{B_1}$. However, by defining  $\phi_{\mathrm{MW}}$ in the simulation, it was found that variation of this angle  has negligible impact. Therefore, it is set to a constant and its effect is neglected in the simulations.

\subsection{DEER Rabi experiments}
\begin{figure}[H]
	\centerline{\includegraphics[scale=0.8]{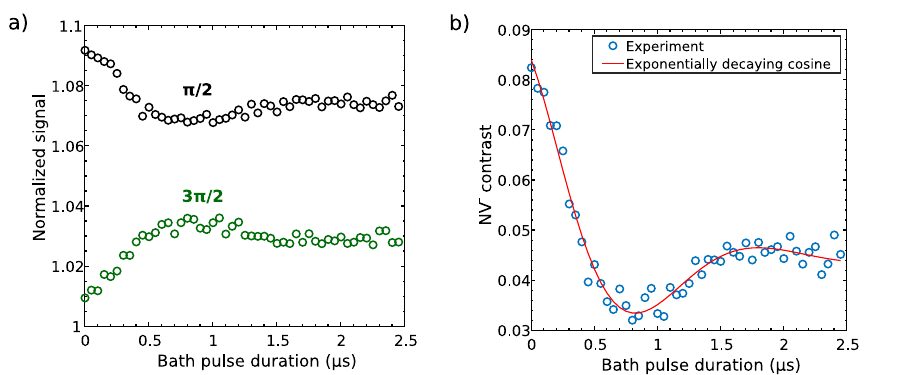}}
	\caption{a) DEER Rabi with a Hahn-$\tau$ of 2\,$\mathrm{\mu}$s driving $^{15}$N$_s^0$ spins at 869.67\,MHz ($B_0$=330\,G). The data in black/green refers to a Hahn-Echo with a $\pi$/2/3$\pi$/2 pulse at the end of the Hahn-Echo. b) NV$^-$ contrast subtracting both signals in a) from each other. The line in red refers to a fit with $A \exp{(\nu t)} \cos{(\Omega t + \delta)} + c$, where $t$ is the pulse duration and $\Omega$ the Rabi frequency. Here, we obtain $\Omega$ = $2\pi$ 0.51\,MHz, and thus $t_{\pi}$=975 ns for the $\pi$-pulse duration.}
	\label{fig:DEERrabi}
\end{figure}

The Rabi frequency $\Omega$ of the bath spins is measured by DEER on an allowed transition of N$_s^0$. For the DEER Rabi experiment the position $T$ of the bath $\pi$-pulse and Hahn-$\tau$ on the NV remains fixed and the bath pulse duration is varied. As throughout of this work, we perform the DEER measurement with a Hahn-sequence mapping the NV-state back onto $m_s$ = 0 ($\pi$/2 pulse at the end) and subtract the data for mapping back into $m_s$ = -1 ($\pi$/2 pulse phase-shifted by 180° at the end). As a consequence, we obtain the absolute spin polarization of the NV center plotted in Figure \ref{fig:DEERrabi}b).

\begin{figure}[H]
	\centerline{\includegraphics[scale=0.8]{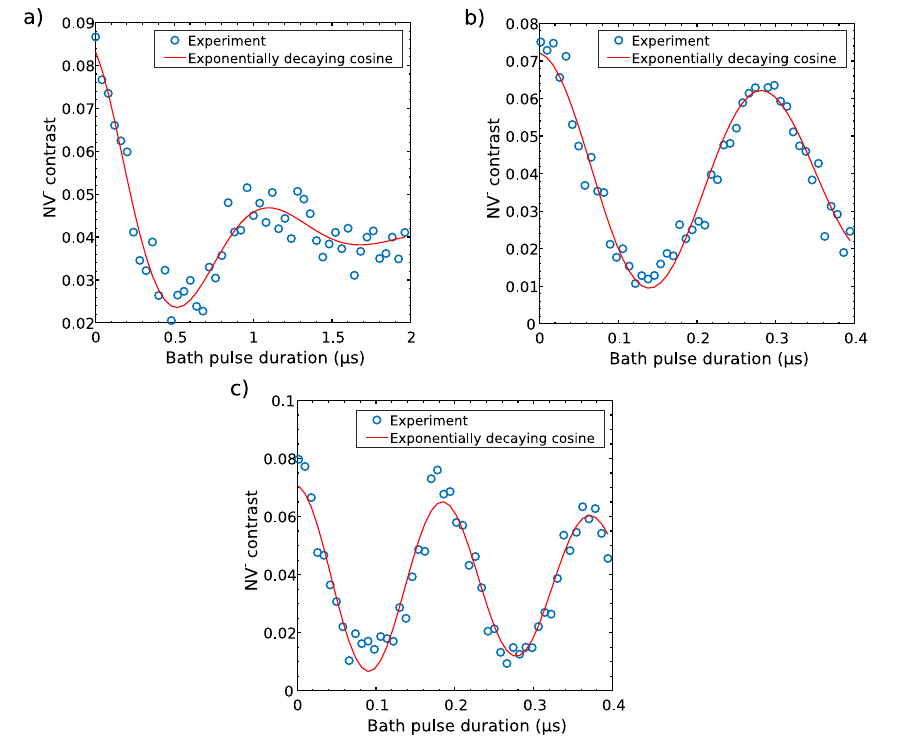}}
	\caption{Representative DEER Rabi experiments driving $^{15}$N$_s^0$ spins at 869.67\,MHz ($B_0=330$\,G) with increasing microwave power from a)-c). The Hahn-$\tau$ on the NV center is 2\,$\mathrm{\mu}$s and for the fitting model refer to the caption of Figure \ref{fig:DEERrabi}. a) $^{15}$N$_s^0$ Rabi frequency $\Omega$ = $2\pi$ 0.86\,MHz, i.e. $t_{\pi}$=580\,ns. b) $\Omega$ = $2\pi$ 3.52 MHz, i.e. $t_{\pi}=142$\,ns. c) $\Omega$ = $2\pi$ 5.35 MHz, i.e. $t_{\pi}=93$\,ns.}
	\label{fig:DEERrabi_representative}
\end{figure}

For the $\pi$-shift experiments to estimate the $^{15}$NVH$^-$ and the $^{15}$N$_s^0$ concentration, we use $\pi$-pulse lengths $t_{\pi}$ of 180\,ns and 280\,ns, respectively. For measuring DEER spectra we decrease the microwave power that we obtain rabi periods of the bath around 1000\,ns and $>$\,1500\,ns in case of high resolution spectra. Figure \ref{fig:DEERrabi_representative} shows representative DEER Rabi experiments for each duration regime.

\subsection{DEER spectra}
In contrast to the main text, we have chosen here to present the spectra not as the NV echo contrast but in form of the normalized DEER contrast $S_{\mathrm{DEER,norm}}$. The representation reminds more of a spectrum obtained from an EPR experiment while the NV echo contrast pronounces the underlying physics in the DEER experiment. The normalized DEER contrast is defined as 

\begin{equation}
S_{\mathrm{DEER,norm}} = 1 - \frac{C_{\mathrm{DEER}}}{C_{0}} , 
\label{eq_normcontrast_EPRlike}
\end{equation}
\noindent where $C_{0}$ is the NV echo contrast without a bath pulse.

\subsubsection{Sample C doped with $^{14}$N}\label{sectionsampleC}
\begin{figure}[H]
	\centerline{\includegraphics[scale=0.8]{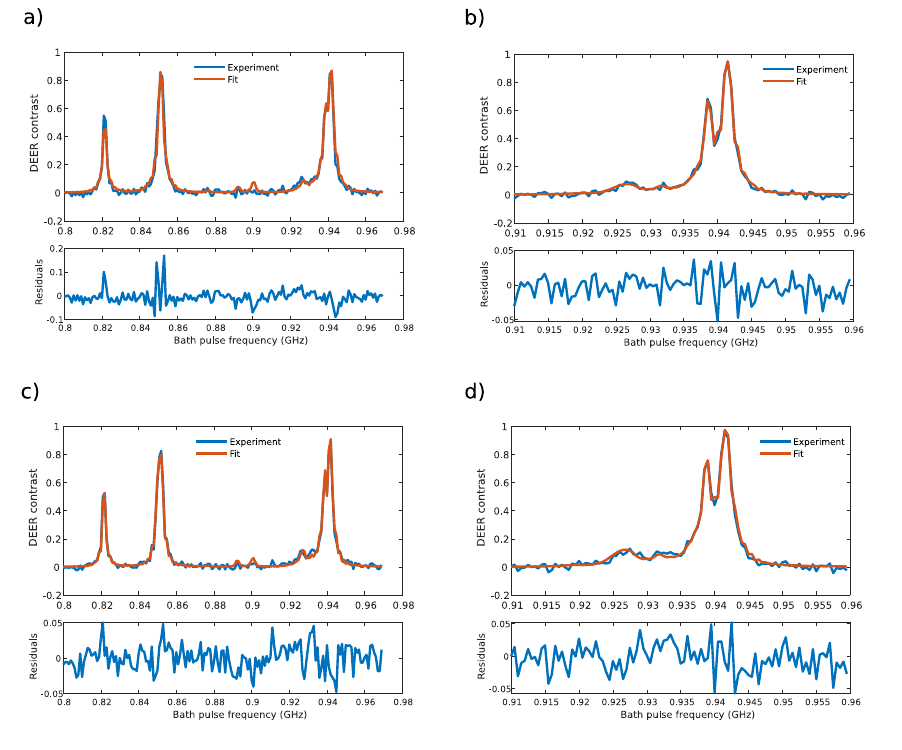}}
	\caption{DEER spectra for Sample C doped with $^{14}$N instead of $^{15}$N. a) Overview spectrum with $t_{\pi}=437$\,ns and Hahn-$\tau$=5 $\mu$s for ensemble C1. b) High resolution spectrum showing the $^{14}$NVH$^-$ region acquired with $t_{\pi}$=907\,ns and a Hahn-$\tau$=9 $\mu$s for ensemble C1. c) Overview spectrum with $t_{\pi}$=445 ns and Hahn-$\tau$=5 $\mu$s for ensemble C2. d) High resolution spectrum showing the $^{14}$NVH$^-$ region acquired with $t_{\pi}=905\,$ns and a Hahn-$\tau=10$\,$\mathrm{\mu}$s for ensemble C2.}
	\label{fig:DEERspec_sampleC}
\end{figure}

\begin{figure}[H]
	\centerline{\includegraphics[scale=0.8]{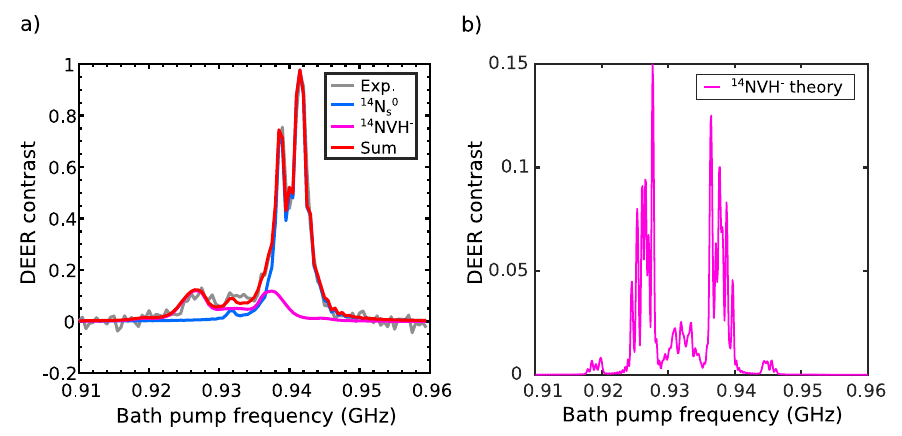}}
	\caption{a) Decomposition of a DEER spectrum for Sample C into $^{14}$NVH$^-$ and $^{14}$N$_s$$^0$ contributions. b) Theoretical DEER spectrum for $^{14}$NVH$^-$ showing the rich, narrow hyperfine structure due to the $I$=1 $^{14}$N nuclear spin.}
	\label{fig:DEERspec_sampleCdecomp}
\end{figure}

Sample C is grown with the same CVD parameters like sample A but using $^{14}$N instead of $^{15}$N. Fig.\ref{fig:DEERspec_sampleC} shows the DEER spectra for ensemble C1 and C2. In Fig.\ref{fig:DEERspec_sampleC}b and d the NVH$^-$ region of the spectrum is shown and the overlap with the allowed $^{14}$N$_s$$^0$ is clearly evident. Fig.\ref{fig:DEERspec_sampleCdecomp}a shows a spectral decomposition of the DEER spectrum in Fig.\ref{fig:DEERspec_sampleC}d and it is clearly visible that the NVH spectrum at higher frequencies entirely overlaps with the $^{14}$N$_s$$^0$ spectrum. The intensity between 922 and 930\,MHz, however, arises purely from NVH$^-$ and can be used to perform a $\pi$-shift experiment (cf. Fig.\ref{fig:DEERpishift_NVH} and Tab. \ref{tab:DEERpishift_NVH}). Fig.\ref{fig:DEERspec_sampleCdecomp}b shows the theoretical DEER spectrum expected for $^{14}$NVH$^-$. The linewidth is sufficiently small to visualize the rich hyperfine structure that we cannot resolve experimentally. The resonances overlap with each other in Fig.\ref{fig:DEERspec_sampleCdecomp}a and we observe the main features in data that is well reproduced taking into account $^{14}$N$_s$$^0$ and $^{14}$NVH$^-$. The quantitative fit method yields a $^{14}$NVH$^-$ concentration for ensemble C2 of (0.37\,$\pm$\,0.02)\,ppm (Tab.\ref{tab:SampleC_fitresults_bath}) what agrees well with the average concentration of Sample A (0.4\,ppm). Likwise, for the linewidth and the N$_s$$^0$ concentration (Tab.\ref{tab:SampleA_fitresults_bath} and Tab.\ref{tab:SampleC_fitresults_bath} we find similar values what makes sense, since the samples A and C are grown with the same parameters (Tab.\ref{tab:CVDparam}). For ensemble C1 we find slightly lower $^{14}$NVH$^-$ concentrations what could arise from a certain heterogeneity in the NVH-distribution, as the fits in Fig.\ref{fig:DEERspec_sampleC}b and d look equally good. For the calculation of the DEER intensity in EasySpin for $^{14}$NVH$^-$ we employ the approach by Glover et. al\cite{Glover.2003} and derive the hyperfine interactions from the ones of $^{15}$NVH$^-$ in Table~\ref{table_Hamiltonian_param_values} by re-scaling according to the gyromagnetic ratios.

Using the $\pi$-shift experiment for ensemble C2 (Tab. \ref{tab:DEERpishift_NVH}) we obtain with (0.26\,$\pm$\,0.01)\,ppm a lower NVH$^-$ concentrations compared with the quantitative fit method (0.37\,$\pm$\,0.02)\,ppm. Although the contribution of $^{14}$N$_s$$^0$ to the mono-exponential decay for ensemble C2 in Fig.\ref{fig:DEERpishift_NVH}a comes mostly from allowed transitions, the explanation for the systematically lower NVH$^-$ concentration from the $\pi$-shift method still applies. The driving of the allowed transition instead of the forbidden explains the significantly faster decay for C2 in Fig. \ref{fig:DEERpishift_NVH}a than for the ensembles in Sample A.

\begin{table}[H]
\caption{\label{tab:SampleC_fitresults_exp}Results from fitting the experimental DEER spectra in Figure \ref{fig:DEERspec_sampleC} with Equation \ref{eq_contrast_2_species_compact1} using EasySpin: experimental parameters Sample C.}
\begin{tabular}{ccccc}
No.& $B_0 (G)$ & $\theta_{\mathrm{MW}}$ ($\deg$) & $\phi_0$ ($\deg$) & $\delta_0$ ($\deg$)\\
\hline
\multicolumn{5}{c}{$\mathrm{N}_s^0 + \mathrm{NVH}^-$: 800\,MHz - 980\,MHz}\\
\hline
C1 & 332.4    &    60\,$\pm$\,8    &    45\,$\pm$\,2  &    55\,$\pm$\,2    \\
C2 & 332.5    &    69\,$\pm$\,6    &    42\,$\pm$\,1  &    54\,$\pm$\,1    \\
\hline
\multicolumn{5}{c}{$\mathrm{NVH}^-$: 910\,MHz - 960\,MHz}\\
\hline
C1 & 332.4    &    48\,$\pm$\,1     &    45\,$\pm$\,14 &    54\,$\pm$\,5    \\
C2 & 332.4    &    58\,$\pm$\,288     &    45\,$\pm$\,9 &    56\,$\pm$\,3    \\
\end{tabular}
\end{table}

\begin{table}[!h]
\caption{\label{tab:SampleC_fitresults_bath}Results from fitting the experimental DEER spectra in Figure \ref{fig:DEERspec_sampleC} with Equation \ref{eq_contrast_2_species_compact1} using EasySpin: spin bath concentration and linewidths for Sample C doped with $^{14}$N.}
\begin{tabular}{ccccc}
No.&   $n_{\mathrm{N}_s^0}$ (ppm) &  $\Delta \omega_{\mathrm{N}_s^0}$ (MHz)   &    $n_{\mathrm{NVH}^-}$ (ppm)   & $\Delta \omega_{\mathrm{NVH}^-}$ (MHz)\\
\hline
\multicolumn{5}{c}{$\mathrm{N}_s^0 + \mathrm{NVH}^-$: 800\,MHz - 980\,MHz}\\
\hline
C1 & 8.05\,$\pm$\,0.18 & 0.88\,$\pm$\,0.05 & 0.23\,$\pm$\,0.04 & 1.70\,$\pm$\,0.72\\
C2 & 8.44\,$\pm$\,0.11 & 0.54\,$\pm$\,0.03 & 0.32\,$\pm$\,0.02 & 1.06\,$\pm$\,0.20\\
\hline
\multicolumn{5}{c}{$\mathrm{NVH}^-$: 910\,MHz - 960\,MHz}\\
\hline
C1 & 6.42\,$\pm$\,0.11 & 0.43\,$\pm$\,0.02 & 0.26\,$\pm$\,0.02 & 1.90\,$\pm$\,0.14\\
C2 & 7.07\,$\pm$\,0.15 & 0.38\,$\pm$\,0.02 & 0.37\,$\pm$\,0.02  & 1.50\,$\pm$\,0.37\\
\end{tabular}
\end{table}

\subsubsection{Further ensembles for Sample A and B}
\begin{figure}[H]
	\centerline{\includegraphics[scale=0.8]{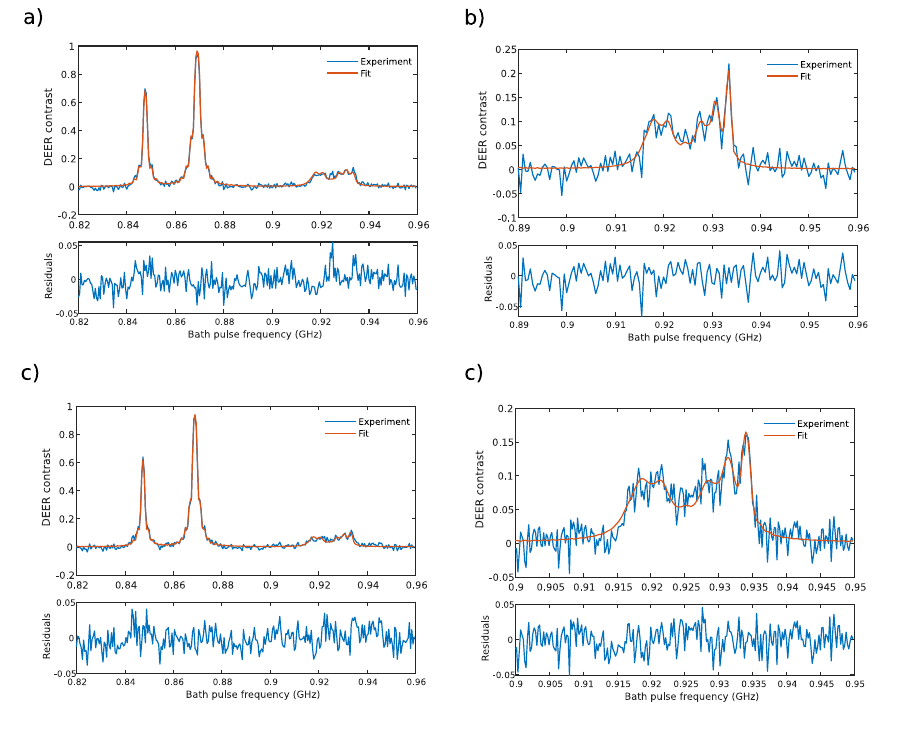}}
	\caption{Additional DEER spectra for Sample A. a) Overview spectrum with $t_{\pi}=530$\,ns and Hahn-$\tau=5$\,$\mu$s for ensemble A2. b) High resolution spectrum showing the $^{15}$NVH$^-$ region acquired with $t_{\pi}=825$\,ns and a Hahn-$\tau=10$\,$\mu$s for ensemble A2. c) Overview spectrum with $t_{\pi}=615$\,ns and Hahn-$\tau=5$\,$\mu$s for ensemble A3. d) High resolution spectrum showing the $^{15}$NVH$^-$ region acquired with $t_{\pi}=740$\,ns and a Hahn-$\tau=10$\,$\mu$s for ensemble A3.}
	\label{fig:DEERspec_sampleA}
\end{figure}

\begin{figure}[H]
	\centerline{\includegraphics[scale=0.8]{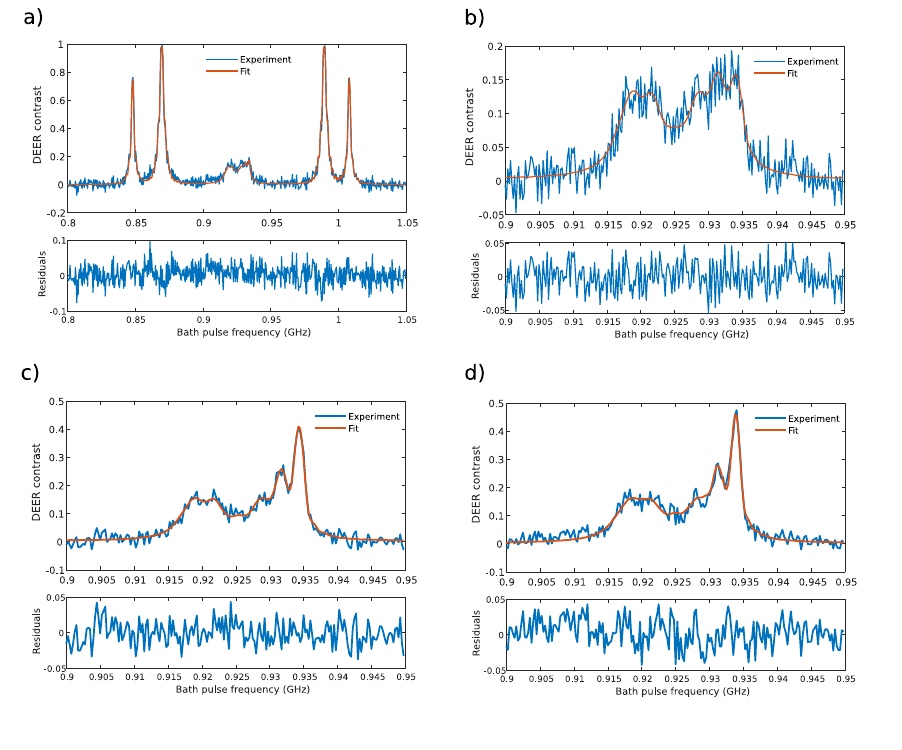}}
	\caption{DEER spectra for Sample B. a) Overview spectrum with $t_{\pi}=537$\,ns and Hahn-$\tau=5$\,$\mu$s for ensemble B1. b) High resolution spectrum showing the $^{15}$NVH$^-$ region acquired with $t_{\pi}=537$\,ns and a Hahn-$\tau=5$\,$\mu$s for ensemble B1. c) High resolution spectrum showing the $^{15}$NVH$^-$ region acquired with $t_{\pi}=665$\,ns and a Hahn-$\tau=7$\,$\mu$s for ensemble B2. d) High resolution spectrum showing the $^{15}$NVH$^-$ region acquired with $t_{\pi}=780$\,ns and a Hahn-$\tau=8$\,$\mu$s for ensemble B3.}
	\label{fig:DEERspec_sampleB}
\end{figure}

In addition to ensemble A1, we have measured another two ensembles, A2 and A3. The corresponding DEER spectra are shown in Fig.\ref{fig:DEERspec_sampleA} while Tab.\ref{tab:SampleA_fitresults_exp} and Tab.\ref{tab:SampleA_fitresults_bath} summarize the fit results. Fig.\ref{fig:DEERspec_sampleB} shows the data for sample B. The full spectrum comprising both the $^{15}$N$^0$$_s$ and the $^{15}$NVH$^-$  resonances is only available for ensemble B1. The pronounced forbidden $^{15}$N$^0$$_s$ transition in Fig.\ref{fig:DEERspec_sampleB}c and d, for ensemble B2 and B3, respectively, arises from significantly smaller angle $\theta_{\mathrm{MW}}$ between Zeeman- and MW-field compared to ensemble B1 (Tab.\ref{tab:SampleB_fitresults_exp}). The angle $\theta_{\mathrm{MW}}$ depends on the alignment of the Zeeman field with respect to the wire used for the application of microwave pulses. The fit results for Sample B can be found in Tab.\ref{tab:SampleB_fitresults_exp} and Tab.\ref{tab:SampleB_fitresults_bath}.

\begin{table}[H]
\caption{\label{tab:SampleA_fitresults_exp}Results from fitting the experimental DEER spectra in the main text and in Figure \ref{fig:DEERspec_sampleA} with Equation \ref{eq_contrast_2_species_compact1} using EasySpin: experimental parameters Sample A}
\begin{tabular}{ccccc}
No.& $B_0 (G)$ & $\theta_{\mathrm{MW}}$ ($\deg$) & $\phi_0$ ($\deg$) & $\delta_0$ ($\deg$)\\
\hline
\multicolumn{5}{c}{$\mathrm{N}_s^0 + \mathrm{NVH}^-$: 820\,MHz - 960\,MHz}\\
\hline
A1 & 329.8 & 65\,$\pm$\,7 & 45\,$\pm$\,51 & 55\,$\pm$\,1 \\
A2 & 329.7 & 60\,$\pm$\,2 & 45\,$\pm$\,1 & 56\,$\pm$\,1 \\
A3 & 329.6 & 54\,$\pm$\,2 & 46\,$\pm$\,1 & 55\,$\pm$\,1 \\
\hline
\multicolumn{5}{c}{$\mathrm{NVH}^-$: 890\,MHz - 960\,MHz}\\
\hline
A1 & 329.9 & 53\,$\pm$\,1 & 45\,$\pm$\,96 & 55\,$\pm$\,9 \\
A2 & 329.7 & 54\,$\pm$\,1 & 45\,$\pm$\,14 & 56\,$\pm$\,3 \\
A3 & 329.9 & 55\,$\pm$\,1 & 45\,$\pm$\,1 & 55\,$\pm$\,7 \\ 
\end{tabular}
\end{table}

\begin{table}[H]
\caption{\label{tab:SampleA_fitresults_bath}Results from fitting the experimental DEER spectra in the main text and in Figure \ref{fig:DEERspec_sampleA} with Equation \ref{eq_contrast_2_species_compact1} using EasySpin: spin bath concentration and linewidths for Sample A}
\begin{tabular}{ccccc}
No.&   $n_{\mathrm{N}_s^0}$ (ppm) &  $\Delta \omega_{\mathrm{N}_s^0}$ (MHz)   &    $n_{\mathrm{NVH}^-}$ (ppm)   & $\Delta \omega_{\mathrm{NVH}^-}$ (MHz)\\
\hline
\multicolumn{5}{c}{$\mathrm{N}_s^0 + \mathrm{NVH}^-$: 820\,MHz - 960\,MHz}\\
\hline
A1 & 7.66\,$\pm$\,0.06 & 0.98\,$\pm$\,0.02 & 0.42\,$\pm$\,0.04 & 1.39\,$\pm$\,0.61\\
A2 & 8.34\,$\pm$\,0.09 & 0.54\,$\pm$\,0.02 & 0.56\,$\pm$\,0.02 & 1.82\,$\pm$\,0.28\\
A3 & 7.21\,$\pm$\,0.08 & 0.52\,$\pm$\,0.02 & 0.42\,$\pm$\,0.02 & 1.70\,$\pm$\,0.23 \\
\hline
\multicolumn{5}{c}{$\mathrm{NVH}^-$: 890\,MHz - 960\,MHz}\\
\hline
A1 & 7.97\,$\pm$\,0.42 & 0.97\,$\pm$\,0.11 & 0.40\,$\pm$\,0.01 & 1.33\,$\pm$\,0.11\\
A2 & 7.16\,$\pm$\,0.61 & 0.14\,$\pm$\,0.11 & 0.43\,$\pm$\,0.02  & 2.10\,$\pm$\,0.21\\
A3 & 6.41\,$\pm$\,0.39 & 0.41\,$\pm$\,0.10 & 0.38\,$\pm$\,0.01  & 2.44\,$\pm$\,0.18\\
\end{tabular}
\end{table}

\begin{table}[H]
\caption{\label{tab:SampleB_fitresults_exp}Results from fitting the experimental DEER spectra in the main text and in Figure \ref{fig:DEERspec_sampleB} with Equation \ref{eq_contrast_2_species_compact1} using EasySpin: experimental parameters Sample B.}
\begin{tabular}{ccccc}
No.& $B_0 (G)$ & $\theta_{\mathrm{MW}}$ $\deg$ & $\phi_0$ $\deg$ & $\delta_0$ $\deg$\\
\hline
\multicolumn{5}{c}{$\mathrm{N}_s^0 + \mathrm{NVH}^-$: 800\,MHz - 1050\,MHz}\\
\hline
B1 & 329.9 & 54\,$\pm$\,2  & 44\,$\pm$\,1  & 55\,$\pm$\,1 \\
\hline
\multicolumn{5}{c}{$\mathrm{NVH}^-$: 900\,MHz - 950\,MHz}\\
\hline
B1 & 329.9 & 52\,$\pm$\,1  & 44\,$\pm$\,20  & 55\,$\pm$\,15 \\
B2 & 330.0 & 39\,$\pm$\,1  & 44\,$\pm$\,13  & 56\,$\pm$\,3 \\
B3 & 329.8 & 29\,$\pm$\,1  & 45\,$\pm$\,14  & 54\,$\pm$\,8 \\ 
\end{tabular}
\end{table}

\begin{table}[H]
\caption{\label{tab:SampleB_fitresults_bath}Results from fitting the experimental DEER spectra in the main text and in Figure \ref{fig:DEERspec_sampleB} with Equation \ref{eq_contrast_2_species_compact1} using EasySpin: spin bath concentration and linewidths for Sample B.}
\begin{tabular}{ccccc}
No.&   $n_{\mathrm{N}_s^0}$ (ppm) &  $\Delta \omega_{\mathrm{N}_s^0}$ (MHz)   &    $n_{\mathrm{NVH}^-}$ (ppm)   & $\Delta \omega_{\mathrm{NVH}^-}$ (MHz)\\
\hline
\multicolumn{5}{c}{$\mathrm{N}_s^0 + \mathrm{NVH}^-$: 800\,MHz - 1050\,MHz}\\
\hline
B1 & 11.02\,$\pm$\,0.14 & 0.66\,$\pm$\,0.02 & 0.97\,$\pm$\,0.04 & 4.90\,$\pm$\,0.33\\
\hline
\multicolumn{5}{c}{$\mathrm{NVH}^-$: 900\,MHz - 950\,MHz}\\
\hline
B1 & 8.90\,$\pm$\,0.71 & 0.90\,$\pm$\,0.25 & 0.81\,$\pm$\,0.02 & 2.44\,$\pm$\,0.18\\
B2 & 10.87\,$\pm$\,0.25 & 0.52\,$\pm$\,0.05 & 0.81\,$\pm$\,0.01  & 2.30\,$\pm$\,0.11\\
B3 & 5.63\,$\pm$\,0.15 & 0.52\,$\pm$\,0.04 & 0.95\,$\pm$\,0.02  & 3.05\,$\pm$\,0.11\\
\end{tabular}
\end{table}

\subsection{$\pi$-shift method}
\begin{figure}[H]
	\centerline{\includegraphics[scale=0.8]{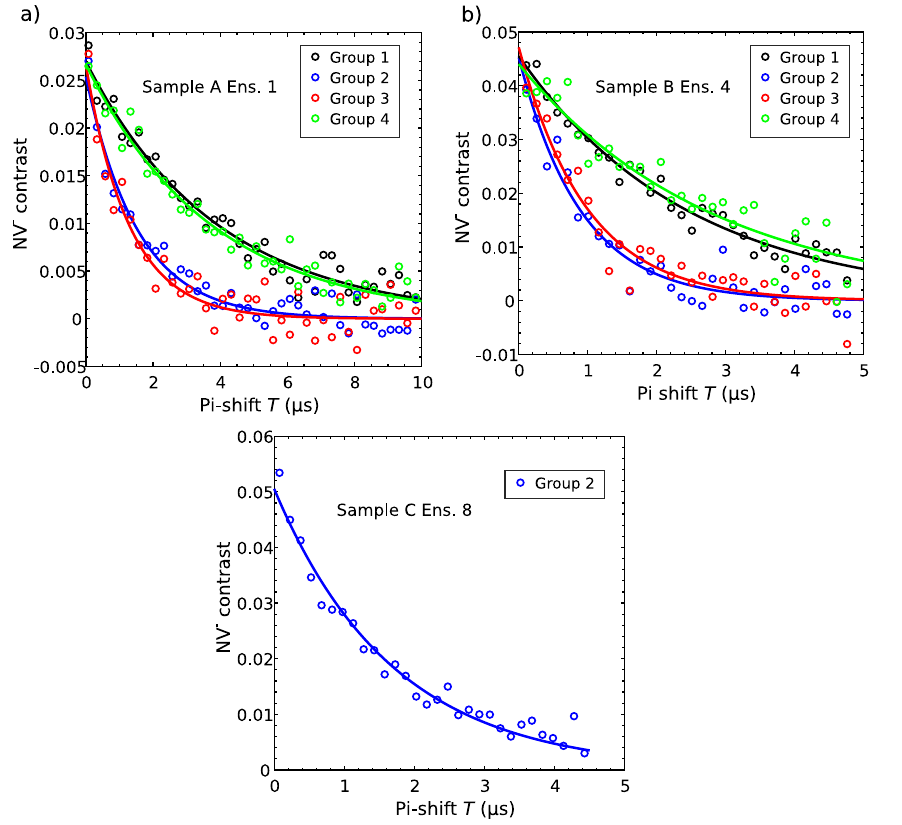}}
	\caption{a)+b) Representative data of the $\pi$-shift experiments on the four $^15$N$_s$$^0$ resonances shown for example in Fig. 1d in the main text. The magnetic field $B_0$ is 330\,G and the solid lines correspond to a fit with a mono-exponential decay of the form $A\exp(-n_{N_s^0} 0.292 \mathrm{\frac{MHz}{ppm}} T)+c$. c) Data of the $\pi$-shift experiments on the $^{14}$N$_s$$^0$ resonance at 851.55\,MHz for ensemble C2 (Sample C) (cf. Figure \ref{fig:DEERspec_sampleC}c). The total concentration of $^{14}$N$_s$$^0$ is obtained by multiplying the concentration obtained from group 2 by 12/3.
	\label{fig:DEERpishift_Ns0}}
\end{figure}

The $\pi$-shift experiments on the N$_s$$^0$ transitions are performed with a bath pulse length of approximately 140\,ns and the data is shown in Fig.\ref{fig:DEERpishift_Ns0}. The corresponding concentrations are summarized in Tab.\ref{tab:DEERpishift_Ns0}. 

\begin{figure}[H]
	\centerline{\includegraphics[scale=0.6]{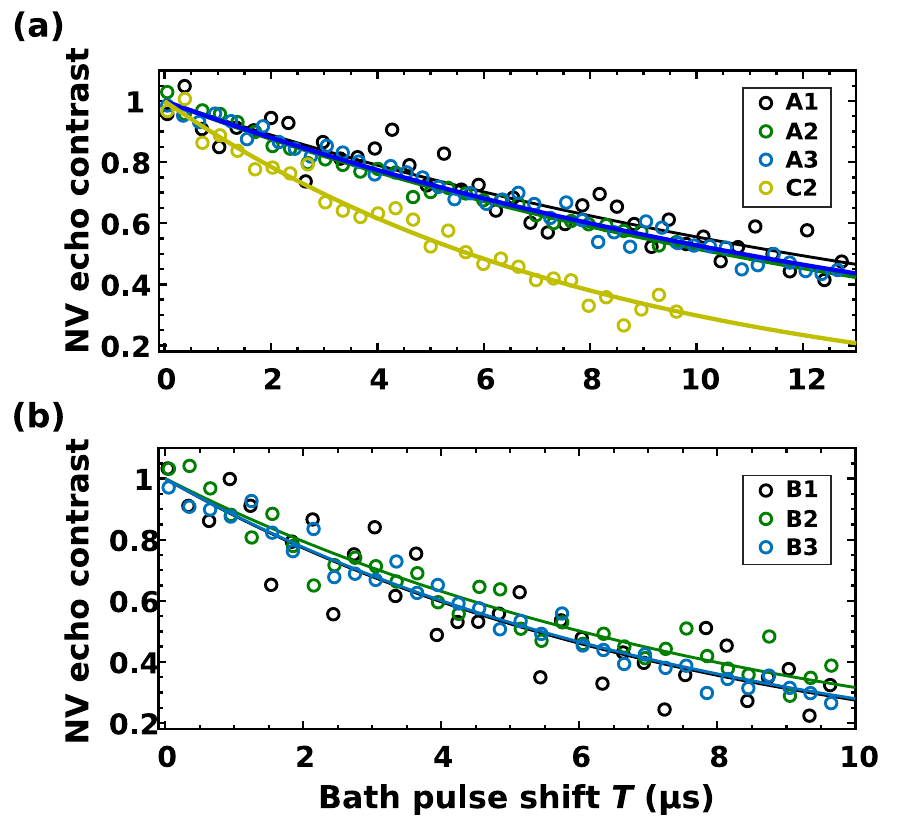}}
	\caption{Experimental results for performing $\pi$-shift experiments on NVH$^-$ resonances at $B_0$ = 330\,G. The solid lines correspond to fits with Equation 3 in the main text considering the fractions of inverted spins from Tab.\ref{tab:DEERpishift_NVH}. (a) NV echo decay curves for Sample A and C. The faster decay for ensemble C2 is related to partially driving of the allowed $^{14}$N$_s$$^0$ transitions around 940\,MHz (Fig.\ref{fig:DEERspec_sampleC}. (b) NV echo decay curves for Sample B.}
	\label{fig:DEERpishift_NVH}
\end{figure}

Fig.\ref{fig:DEERpishift_NVH} shows the NV echo contrast for performing a $\pi$-shift experiment on the NVH$^-$ frequencies provided in Tab.\ref{tab:DEERpishift_NVH}. The decay when driving NVH$^-$ is significantly longer than for N$_s$$^0$ in Fig.\ref{fig:DEERpishift_Ns0} since the related concentration is by one order of magnitude smaller. Regarding the relatively fast decay observed for C2 in Fig.\ref{fig:DEERpishift_Ns0}a, refer to the discussion in section \ref{sectionsampleC}. The calculated fractions of inverted spins and the NVH$^-$ concentrations from the fit are presented in Tab.\ref{tab:DEERpishift_NVH}.

\begin{table}[H]
\caption{\label{tab:DEERpishift_Ns0}Total concentration of N$_s$$^0$ estimated by performing a $\pi$-shift DEER experiment on each of the four $^{15}$N$_s$$^0$ transitions and adding the obtained concentrations (Sample A and B), see also Figure \ref{fig:DEERpishift_Ns0} a) + b). For Sample C the measurement is performed only for the resonance at 851.55\,MHz (Figure \ref{fig:DEERspec_sampleC}) and then multiplied by its spectral weight $\frac{12}{3}$ (Figure \ref{fig:DEERpishift_Ns0} c)).}
\begin{tabular}{cc}
No.& $n_{\mathrm{N}_s^0}$ (ppm)\\
\hline
\multicolumn{2}{c}{Sample A}\\
\hline
A1 & 6.7 $\pm$ 0.6  \\
A2 & 8.4 $\pm$ 0.3  \\
A3 & 7.4 $\pm$ 0.2  \\
\hline
\multicolumn{2}{c}{Sample B}\\
\hline
B1 & 9.9 $\pm$ 0.9  \\
B2 & 10.2 $\pm$ 0.9  \\
B3 & 9.9 $\pm$ 0.6  \\
\hline
\multicolumn{2}{c}{Sample C}\\
\hline
C1 & -  \\
C2 & 8.1 $\pm$ 0.7  \\
\end{tabular}
\end{table}

\begin{table}[H]
\caption{\label{tab:DEERpishift_NVH}Total concentration of NVH$^-$ ($n_{\mathrm{NVH}^-}$)  estimated by performing a $\pi$-shift DEER experiment driving $^{15}$NVH$_s$$^0$ transitions at the frequency $f_\pi$ along with the inverted spin fractions $x_{\mathrm{N}_s^0}$ and $x_{\mathrm{NVH}^-}$. The magnetic field is set to 330\,G.}
\begin{tabular}{ccccc}
No.&  $f_\pi$ (MHz)  &  $x_{\mathrm{NVH}^-}$  &  $x_{\mathrm{N}_s^0}$  &    $n_{\mathrm{NVH}^-}$ (ppm)  \\
\hline
\multicolumn{5}{c}{Sample A}\\
\hline
A1 & 919.2 & 0.448 & 0.007 & 0.34 $\pm$ 0.07  \\
A2 & 919.0 & 0.468 & 0.007 & 0.35 $\pm$ 0.03  \\
A3 & 920.0 & 0.467 & 0.006 & 0.37 $\pm$ 0.03  \\
\hline
\multicolumn{5}{c}{Sample B}\\
\hline
B1 & 920.0 & 0.597 & 0.004 & 0.67 $\pm$ 0.14  \\
B2 & 920.0 & 0.436 & 0.008 & 0.72 $\pm$ 0.11  \\
B3 & 920.0 & 0.395 & 0.008 & 0.90 $\pm$ 0.09  \\
\hline
\multicolumn{5}{c}{Sample C}\\
\hline
C2 & 927.0 & 0.549 & 0.034 & 0.257 $\pm$ 0.10  \\  \\
\end{tabular}
\end{table}

\subsection{Nitrogen concentration range for successful NVH detection}
In this work, we investigated as-grown samples of nitrogen concentration in between 13.8\,ppm and 16.7\,ppm. However, the DEER methods that we demonstrated seem capable of serving for investigating samples with different composition, one adaptation being with regards to the value of the pulse spacing  $\tau$ on the  NV$^-$.  In general, it appears that a good choice for $\tau$ is, given by $2\tau = T_{2,\mathrm{NV}}$, where  $T_{2,\mathrm{NV}}$ is the decoherence time of the NV$^-$ in the Hahn-echo.  A study on samples in which substitutional nitrogen is the dominant impurity, reported in Bauch et al.~\cite{Bauch.2020}, established that in the range of nitrogen concentrations \mbox{0.5 ppm $ \lesssim \mathrm{[N]} <$ 300 ppm, $T_{2,\mathrm{NV}}$} is mostly determined by the nitrogen content through the scaling law $T_{2,\mathrm{NV}}^{-1} = a \mathrm{[N]}$ with $a=160\pm12$~\si{\micro\second}~ppm.  Therefore, for samples that have lower concentration of nitrogen (but still $\gtrsim 0.5$ ppm), the possibility to define longer $\tau$ without important loss of signal on the NV$^-$ is beneficial, as it may be used to successfully detect \NVH{} and other defects despite their lower expected concentration. For instance, at a [N]=1 ppm, $T_2 \sim 160$~\si{\micro\second} implies that $\tau = T_2 /2 = 80$~\si{\micro\second} can be set, which is eight times higher than the longest $\tau$ we used in this study, and would result in a sensitivity improvement. Below $\sim0.5$ ppm, still according to the work from Bauch et al.~\cite{Bauch.2020}, the decoherence of NV$^-$ becomes gradually decoupled from the nitrogen. At low densities, it reaches a limit ($T_{2,\mathrm{NV}} \sim 700$~\si{\micro \second}) so that the sensitivity of DEER techniques will not improve anymore - in fact, it will at one point decrease, assuming a constant NV$^-$:N ratio, due to the lower amount of NV$^-$ in a given optical detection volume. Another regime would correspond to nitrogen-rich samples, with [N]$\gtrsim 20$ ppm, there, one would need probably to decrease the $\tau$ value to adapt to the expected shorter $T_{2,\mathrm{NV}}$. Although we believe the experiments we demonstrated could be reproduced in such samples, new difficulties might arise, as we expect, mainly  from deteriorating spectral linewidths. For instance, in $^{15}$N samples, the $^{15}$N-related hyperfine structure of \NVH{} will become gradually harder to resolve with the neighbouring lines merging together, as a consequence of the higher nitrogen content. Rigorous attribution of the signal to \NVH{}  might then require using further advanced techniques (in combination with DEER) such as electron-nuclear double resonance (ENDOR).~\cite{Baker.1994} If \NVH{} can be assumed to be an impurity, with concentration far above that of defects with neighbouring spectral signature (VH$^0$), we speculate that the techniques we described could be used to quantify precisely the \NVH{} content, despite the low spectral resolution, in nitrogen-rich samples up to $\sim100$\,ppm. 

\clearpage
\pagebreak

\bibliographystyle{unsrt}

\clearpage
\pagebreak

\end{document}